\documentclass[10pt]{article}
\usepackage{ametsoc2col}
\newcommand{\myabstract}{New parameterizations for the spectral dissipation of wind-generated waves 
are proposed. The rates of dissipation have no predetermined spectral shapes
and are functions of the wave spectrum, in a 
way consistent with observation of wave breaking and swell dissipation 
properties. Namely, swell dissipation is nonlinear and proportional to 
the swell steepness, and wave breaking only affects spectral components such that the non-dimensional spectrum exceeds the threshold at which waves are 
observed to start breaking. An additional source of short wave dissipation 
due to long wave breaking is introduced, together with a reduction of  
wind-wave generation term for short waves, otherwise taken from Janssen (J. Phys. 
Oceanogr. 1991). These parameterizations are combined and calibrated with the 
Discrete Interaction Approximation of Hasselmann et al. (J. Phys. Oceangr. 
1985) for the nonlinear interactions. Parameters are  adjusted to reproduce 
observed shapes of directional wave spectra, and the variability of spectral 
moments with wind speed and wave height. The wave energy balance is verified 
in a wide range of conditions and scales, from the global ocean to coastal  settings. Wave height, peak and 
mean periods, and spectral data are validated using in situ and remote sensing 
data. Some systematic defects are still present, but the parameterizations 
probably yield the most accurate overall estimate of wave parameters to date. Perspectives for further improvement are 
also given.}
\begin{document}
%
%
\title{\textbf{\large{Semi-empirical dissipation source functions for ocean waves: \\
 Part I, definition, calibration and validation.}}}
%
%
\author{\textsc{Fabrice Ardhuin}
				\thanks{\textit{Corresponding author address:} 
				Fabrice Ardhuin, 13 rue du Chatellier, 
				29603 BREST Cedex, France. 
				\newline{E-mail: ardhuin@shom.fr}, }, \textsc{Jean-Fran{\c c}ois Filipot and Rudy Magne}\\
\textit{\footnotesize{Service Hydrographique et Oc{\'e}anographique de la Marine, Brest, France}}
\and
\centerline{\textsc{Erick Rogers}}\\
\centerline{\textit{\footnotesize{Oceanography Division, Naval Research Laboratory, Stennis Space Center, MS, USA}}}
\and
\centerline{\textsc{Alexander Babanin}}\\
\centerline{\textit{\footnotesize{Swinburne University, Hawthorn, VA, Australia}}}
\and
\centerline{\textsc{Pierre Queffeulou}}\\
\centerline{\textit{\footnotesize{Ifremer, Laboratoire d'Oc{\'e}anographie Spatiale, Plouzan{\'e}, France}}}
\and
\centerline{\textsc{Lotfi Aouf and Jean-Michel Lefevre}}\\
\centerline{\textit{\footnotesize{UMR GAME, M{\'e}t{\'e}o-France - CNRS, Toulouse, France}}}
\and
\centerline{\textsc{Aron Roland}}\\
\centerline{\textit{\footnotesize{Technological University of Darmstadt, Germany}}}
\and
\centerline{\textsc{Andre van der Westhuysen}}\\
\centerline{\textit{\footnotesize{Deltares, Delft, The Netherlands}}}
\and
\centerline{\textsc{Fabrice Collard}}\\
\centerline{\textit{\footnotesize{CLS, Division Radar, Plouzan{\'e}, France}}}
}

%
\ifthenelse{\boolean{dc}}
{
\twocolumn[
\begin{@twocolumnfalse}
\amstitle

\begin{center}
\begin{minipage}{13.0cm}
\begin{abstract}
	\myabstract
	\newline
	\begin{center}
		\rule{38mm}{0.2mm}
	\end{center}
\end{abstract}
\end{minipage}
\end{center}
\end{@twocolumnfalse}
]
}
{
\amstitle
\begin{abstract}
\myabstract
\end{abstract}
\newpage
}
%
\section{Introduction}

\subsection{On phase-averaged models}
Spectral wave modelling has been performed for the last 50 years, using the
wave energy balance equation \citep{Gelci&al.1957}.  This model for the evolution of spectral densities of the surface elevation variance $F$
distributed over frequencies $f$ and directions $\theta$ can be put in the form
\begin{equation}
\frac{{\mathrm d} F(f,\theta)}{{\mathrm d} t}= S_{\mathrm{atm}}(f,\theta) +
S_{\mathrm{nl}}(f,\theta) +S_{\mathrm{oc}}(f,\theta)+S_{\mathrm{bt}}(f,\theta),\label{Energy_balance}
\end{equation}
where the Lagrangian derivative is the rate of change of the spectral density
when following a wave packet at its group speed in physical and spectral space.
This spectral advection particularly includes changes in direction due to the Earth sphericity and refraction over varying topography \citep[e.g.][]{Munk&Traylor1947,Magne&al.2007} and currents, and changes in wavelength or period in similar conditions \citep{Barber1949}. 

The source functions on the right hand side are separated into an atmospheric
source function $S_{\mathrm{atm}}$, a nonlinear scattering term
$S_{\mathrm{nl}}$, an ocean source $S_{\mathrm{oc}}$, and a
bottom source $S_{\mathrm{bt}}$. 

This separation, like any other, is largely arbitrary. For example, waves that break are highly nonlinear and thus the effect of breaking waves that is contained in $S_{\mathrm{oc}}$ is intrisically related to the non-linear evolution term contained in $S_{\mathrm{nl}}$. Yet, 
compared to the usual separation of deep-water evolution into wind input, non-linear
interactions, and dissipation, it has the benefit of identifying where the
energy and momentum is going to or coming from, which is a necessary feature when ocean waves are used to drive or are coupled with atmospheric or ocean circulation models \citep[e.g.][]{Janssen&al.2004, Ardhuin&al.2008}. 

$S_{\mathrm{atm}}$, which gives
the flux of energy from the atmospheric non-wave motion to the wave motion, is
the sum of a wave generation term $S_{\mathrm{in}}$ and a wind generation term
$S_{\mathrm{out}}$ (often referred to as ``negative wind input'', i.e. a wind output). The nonlinear scattering term $S_{\mathrm{nl}}$
represents all processes that lead to an exchange of wave energy and momentum between the
different spectral components. In deep and intermediate water depth, this is
dominated by cubic interactions between quadruplets of wave trains, while quadratic
nonlinearities play an important role in shallow water \citep[e.g.][]{WISE2007}. The ocean
source $S_{\mathrm{oc}}$ may accomodate wave-current interactions\footnote{In the presence of variable current, the 
source of energy for the wave field, i.e. the work of the radiation stresses, is generally hidden when the energy balance is written as an action balance (e.g. Komen et al. 1994).} and
interactions of surface and internal waves, but it will be here restricted to
wave breaking and wave-turbulence interactions.

 The basic principle
underlying equation (\ref{Energy_balance}) is that waves essentially propagate as a superposition
of almost linear wave groups that evolve on longer time scales as a result of weak-in-the-mean processes \citep[e.g.][]{WAMBook}. Recent reviews have questioned the possibility of further improving numerical
wave models without changing this basic principle \citep{Cavaleri2006}.
Although this may be true in the long term, we demonstrate here that it is
possible to improve model results significantly by including more
physical features in the source term parameterizations. The main advance
proposed in the present paper is the adjustment of a shape-free dissipation
function based on today's empirical knowledge on the breaking of random waves \citep{Banner&al.2000,Babanin&al.2001} and the dissipation of swells over long
distances \citep{Ardhuin&al.2009}. The present formulations are still semi-empirical, in the sense that they are not based on a detailed physical model of dissipation processes, but they demonstrate that progress is possible. This effort opens the way for completely physical parameterizations \citep[e.g.][]{Filipot&al.2010} that will eventually provide new applications for wave models, such as the estimation of statistical parameters for breaking waves, including whitecap coverage and foam thickness. Other efforts, less empirical in nature, are also under way to arrive at better parameterizations \citep[e.g.][]{Banner&Morison2006,Babanin&al.2007b, Tsagareli2008}, but they yet have to produce a practical alternative for wave forecasting and hindcasting.

\subsection{Shortcomings of existing parameterizations}
All wave dissipation parameterizations up to the work of \cite{vanderWesthuysen&al.2007} had no quantitative relationship with observed features of wave dissipation, and the parameterizations were generally used as set of tuning knobs to close the wave energy balance. The parameterization of the form proposed by \cite{Komen&al.1984} have produced a family loosely justified by the so-called `random pulse' theory of \citep{Hasselmann1974}. These take a generic form
\begin{equation}
S_{\mathrm{oc}} \left(f,\theta\right)=C_{\mathrm{ds}} g^{0.5} k_{r}^{4.5} H_s^4 \left[\delta_1 \frac{k}{k_r} + \delta_2 \left(\frac{k}{k_r}\right)^2\right]\label{WAM4dis},
\end{equation}
in which $C_{\mathrm{ds}}$ is a negative constant, and $k_{r}$ is an energy-weighted mean wavenumber defined from the entire spectrum, and $H_s$ is the significant wave height. In the early and latest parameterizations, the following definition was used 
\begin{equation}
k_r = \left[\frac{16}{H_s^2}\int_0^{f_{\max}} \int_0^{2 \pi} k^r E \left(f,\theta\right) {\mathrm d}f {\mathrm d} \theta\right]^{1/r},
\end{equation}
where $r$ is a chosen real constant, typically $r=-0.5$ or $r=0.5$.

These parameterizations are still widely used in spite of inconsistencies in the underlying theory. Indeed, if whitecaps do act as random pressure pulses, their average work on the underlying waves only occurs because of a phase correlation between the vertical orbital velocity field and the moving whitecap position, which travels with the breaking wave. In reality the horizontal shear is likely the dominant mechanism \citep{Longuet-Higgins&Turner1974}, but the question of correlation remains the same. For any given whitecap, such a correlation cannot exist for all spectral wave components: a whitecap that travels with one wave leads to the dissipation of spectral wave components that propagate in similar directions, with comparable phase velocities. However, whitecaps moving in one direction will give (on average) a zero correlation for waves propagating in the opposite direction because the position of the crests of these opposing waves  are completely random with respect to the whitecap position. As a result, not all wave components are dissipated by a given whitecap (others should even be generated), and the dissipation function cannot take the spectral form given by \cite{Komen&al.1984}. 

A strict interpretation of the pressure pulse model gives a zero dissipation for swells in the open ocean because the swell wave phases are uncorrelated to those of the shorter breaking waves. There is  only a negligible dissipation due to short wave modulations by swells and preferential breaking on the swell crests \citep{Phillips1963,Hasselmann1971,Ardhuin&Jenkins2005}.
Still, the \cite{Komen&al.1984} type dissipation terms are applied to the entire spectrum, including swells, without any physical justification. 

In spite of its successful use for the estimation of the significant wave height $H_s$ and peak period $T_p$, these
fixed-shape dissipation functions, from Komen et al. (1984) up to \cite{Bidlot&al.2007},  have built-in defects. Most conspicuous is the
spurious amplification of wind sea growth in the presence of swell \citep[e.g.][]{vanVledder&Hurdle2002}, which is contrary to all observations \citep{Dobson&al.1989,Violante-Carvalho&al.2004,Ardhuin&al.2007}. Associated wih that defect also
comes an underestimation of the energy level in the inertial range, making these
wave models ill-suited for remote sensing applications, as will be exposed below. 

Also, these parameterizations typically give a decreasing dissipation of swell with increasing swell steepness, contrary to all observations from \cite{Darbyshire1958} to \cite{Ardhuin&al.2009}. This effect is easily seen by taking a sea state composed of a swell and wind sea of energy $E_1$ and $E_2$ and mean wavenumbers $k_1$ and $k_2$, respectively, with $k_2 > k_1$. The overall mean wavenumber is 
\begin{equation}
k_r=\left[\left(k^r_1 E_1 + k^r_2 E_2\right)/(E_1+E_2)\right]^{1/r}.
\end{equation}
Equation (\ref{WAM4dis}) gives a  dissipation that is proportional to $k_r^{3.5}(E_1+E_2)$ in the low frequency limit. Now, if we keep $k_1$, $k_2$ an $E_2$ constant and only increase the swell energy $E_1$, the relative change in dissipation is, according to (\ref{WAM4dis}), proportional to $x=3.5[(k_1/k_r)^r-1]/r+2$. For $r=0.5$, as used by \cite[][hereinafter BAJ]{Bidlot&al.2005}, and $k_1/k_r <0.51$, $x$ is negative (i.e. the dissipation decreases with increasing swell energy). For equal energy in sea and swell, this occurs when $k_1/k_2 < 0.3$, which is generally the case with sea and swell in the ocean. This erroneous decrease of swell dissipation with increasing swell steepness is reduced when the model frequency range is limited to maximum frequency of 0.4~Hz, in which case the lowest winds (less than 5 m/s) are unable to produce a realistic wind sea level, hence limiting the value of $k_r$ to relatively small values.

An alternative and widely used
formulation has been proposed by \cite{Tolman&Chalikov1996}, and some of its features are worth noting. It 
combines two distinct dissipation formulations for high and low frequencies, with a transition at two times the wind sea peak frequency. Whereas \cite{Janssen&al.1994} introduced the use of two terms, $k$ and $k^2$  in eq. (\ref{WAM4dis}), in order to match the very different balances in high and low frequency parts of the spectrum, they still had a common fixed coefficient, $C_{\mathrm{ds}} g^{0.5} k_{r}^{4.5} H_s^4$. In \cite{Tolman&Chalikov1996} these two dissipation terms are completely distinct, the low frequency part being linear in the spectrum and proportional to wind friction velocity $u_\star$, the high frequency part is also linear and proportional to  $u_\star^2$. In this formulation the frequency dependence of the two terms is also prescribed. \cite{Tolman&Chalikov1996}  further included swell attenuation by the wind, based on numerical simulations of the airflow above waves \citep{Chalikov&Belevich1993}, here noted $S_{\mathrm{out}}$.  At relatively short fetches, these source terms are typically a factor 2 to 3 smaller than those of \cite{Janssen&al.1994}, which was found to produce important biases in wave growth and wave directions at short fetch \citep{Ardhuin&al.2007}. Another set of parameterizations was proposed  by \cite{Makin&Stam2003}. It is appropriate for high winds conditions but does not produce accurate results in moderate sea states \citep{Lefevre&al.2004}. Finally, among the many formulations proposed we may cite one by \cite{Polnikov&Inocentini2008}, but its accuracy  appears generally less than with the model presented here, in particular for mean periods. 

Based on observations of large wave height gradients in rapidly varying currents, \cite{Phillips1984} proposed a dissipation rate proportional to the non-dimensional spectrum $B$, also termed 'saturation spectrum'.  \cite{Banner&al.2000}  indeed found a correlation between the breaking probability of dominant waves and the saturation, when the latter is integrated over a finite frequency bandwidth and all directions., with breaking occuring when $B$ exceeds a threshold $B_r$.  \cite{Alves&Banner2003} proposed to define the dissipation $S_{\mathrm{oc}}$ by $B/B_r$ to some power, multiplied by a Komen-type dissipation term. Although
 this approach avoided the investigation of the dissipation of non-breaking waves, it imported all the above mentionned
 defects of that parameterization. Further, these authors used a value for $B_r$ that is much higher than suggested by
 observations, which tends to disconnect the parameterization from the observed effects
 \citep{Babanin&vanderWesthuysen2008}.

The use of a saturation parameter was taken up again by
 \cite{vanderWesthuysen&al.2007}, hereinafter WZB, who, Like Alves and Banner (2003), integrated the saturation spectrum over directions, giving
\begin{equation}
B\left(f\right)=
\int_{0}^{2 \pi} k^3  F(f,\theta^{\prime}) C_g / (2 \pi ) \mathrm d
\theta^{\prime} \label{defB}.
\end{equation}
From this, they defined the source function
\begin{equation}
S_{\mathrm{oc,WZB}}\left(f,\theta\right)=-C \sqrt{gk} \left[\frac{B(f)}{B_r}\right]^{p/2},
\end{equation}
where $C$ is a positive constant, $B_r$ is a constant saturation threshold and and $p$ is a coefficient that varies both with the wind friction velocity $u_\star$ and the degree of saturation $B(f)/B_r$ with, in particular, $p \approx 0$ for $B(f)<0.8 B_r$. 
For non breaking waves, when $p\approx 0$, the dissipation is too large by at least one order of magnitude, making the parameterization unfit for oceanic scale applications with wave heights in the Atlantic underpredicted by about 50\% \citep{Ardhuin&LeBoyer2006}. In \cite{vanderWesthuysen2007} this was addressed 
by reverting back to Komen et al. (1984) dissipation for 
non-breaking waves, but no solution for the 
dissipation of these spectral components was proposed.
In WZB, the increase of $p$ with the inverse wave age $u_\star/C$ was designed to increase $S_{\mathrm{oc}}$ at high frequency, which was needed to obtain a balance with the $S_{\mathrm{atm}}$ term in equation (\ref{Energy_balance}). This indicates that, besides the value of the saturation $B_r$, other factors may be important, such as the directionality of the waves \citep{Banner&al.2002}. Other observations clearly show that the breaking rate of high frequency waves is much higher for a given value of $B$, probably due to cumulative effects by which the longer waves are modifying the dissipation of shorter waves. 

\cite{Banner&al.1989} and \cite{Melville&al.2002} have shown how breaking waves suppress the short waves on the surface, and we will show here that a simple estimation of the dominant breaking rates based on the observations by \cite{Banner&al.2000} suggests that this effect is dominant for wave frequencies above three times the windsea peak frequency. \cite{Young&Babanin2006b} arrived at the same conclusion from the examination of wave spectra, and proposed a parameterization for $S_{\mathrm{oc}}$ that included a new term, the cumulative term,  to represent theis effect. Yet, their estimate was derived for very strong wind-forcing conditions only. Further, 
their interpretation of the differences in parts of a wave record with breaking and non-breaking waves implies an underestimation of the dissipation rates because the breaking waves have already lost some energy when they are observed and the non-breaking waves are not going to break right after they have been observed.
Also, since the spectra are different, nonlinear interactions must be different, even on this relatively small time scale \citep[e.g.][figure 5]{Young&vanVledder1993}, and the differences in spectra may not be the result of dissipation alone. 


Finally, the recent measurement of swell dissipation by \cite{Ardhuin&al.2009b} has revealed that the dissipation of non-breaking waves is essentially a function of the wave steepness, and a very important process for ocean basins larger than 1000~km. Because of the differences between coastal
and larger scale sea states \citep[e.g.][]{Long&Resio2007}, it is paramount to
verify the source function parameterizations at all scales, in order to provide a
robust and comprehensive parameterization of wave dissipation.

\subsection{A new set of parameterizations and adjustments to get adequate balances}
It is thus time to combine the existing knowledge on the dissipation of breaking and non-breaking waves to provide 
an improved parameterization for the dissipation of waves. Our objective is to provide a robust parameterization that improves existing wave models. For this we will use the parameterization by BAJ as a benchmark, because it was shown to provide the best forecasts on global scales \citep{Bidlot&al.2007b} before the advent of the parameterizations presented here. BAJ is also fairly close to the widely used ``WAM-Cycle 4'' parameterization by Janssen and others \citep{WAMBook}.

We will first present a general form of the dissipation terms based on observed wave dissipation features. The degrees of freedom in the parameterization are then used to adjust the model result. In particular we adujst a cumulative breaking effect and a wind sheltering effect that, respectively, dissipates and reduces the wind input to short waves as a function of longer waves characteristics. A comprehensive validation of wave parameters is then presented using field experiments and a one year hindcast of waves at the global and regional scale, in which all possible wave measurements are considered, with significant wave heights ranging from 0 to 17~m. The model is further validated with independent data at regional and global scales. 

Tests and verification in the presence of currents, and using a more realistic parameterizations of wave-wave interactions will be presented in parts II and III. These may also include some replacement of the arbitrary choices made here in the details of the dissipation parameters, with physically-motivated expressions.

\section{Parameterizations}
Several results will be presented, obtained by a numerical integration of the energy balance. Because numerical choices can have  important effects \citep[e.g.][]{Tolman1992b,Hargreaves&Annan2001}, a few details should be given. All calculations are performed with the WAVEWATCH III$^{\rm{TM}}$ modelling framework \citep{Tolman2008,Tolman2009}, hereinafter WWATCH, using the third order spatial and spectral advection scheme, and including modifications of the source terms described here.
In all cases ran with WWATCH, the source terms are integrated with the fully implicit scheme of \cite{Hargreaves&Annan2001},  combined  with the adaptative time step and limiter method of \cite{Tolman2002b}, in which a  minimum time step of 10~s is used, so that the limiter on wind-wave growth is almost never activated. The diagnostic tail, proportional to $f^{-5}$  is only imposed at a cut-off frequency $f_c$ set to 
\begin{equation}
f_c=f_{\mathrm{FM}}  f_m. \label{def_fc}
\end{equation}
Here we take $f_{\mathrm{FM}}=10$ and define the mean frequency as $f_m$=$1/T_{m0,1}$. Hence $f_c$ is generally above the maximum model frequency that we fixed at 0.72~Hz, and the high frequency tail is left to evolve freely. Some comparison tests are also done with other parameterizations using a lower value of $f_c$, typically set at 2.5  $f_m$ \citep{Bidlot&al.2007}. In such calculations, although the net source term may be non-zero at frequencies above $f_c$, there is no spectral evolution due to the imposed tail. 

\subsection{Nonlinear wave wave interactions}
All the results discussed and presented in this section are obtained with the Discrete Interaction Approximation of \cite{Hasselmann&al.1985b}. The coupling coefficient that gives the magnitude of the interactions is $C_{\mathrm{nl}}$. Based on comparisons with exact calculations,  \cite{Komen&al.1984} adjusted  the value of $C_{\mathrm{nl}}$ to  $2.78\times10^7$, which is the value used by \cite{Bidlot&al.2005}. Here this constant will be allowed to vary slightly. This parameterization is well known for its shortcomings \citep{Banner&Young1994}, and the adjustment of other parameters probably compensates for some of these errors. This matter will be fully discussed in Part III.  

\subsection{Swell dissipation}
Observations of swell dissipation are consistent with the effect of friction at the air-sea interface \citep{Ardhuin&al.2009b}, resulting in a flux of momentum from the wave field to the wind \citep{Harris1966}. 
We thus write the swell dissipation as a negative contribution $S_{\mathrm{out}}$ which is added to $S_{\mathrm{in}}$ to make the wind-wave source term 
$S_{\mathrm{atm}}$. 

Using the method of \cite{Collard&al.2009}, a systematic analysis of swell observations by \cite{Ardhuin&al.2009b} showed that the swell dissipation is non-linear, possibly related to a laminar-to-turbulent transition of the oscillatory boundary layer over swells. 
Defining the boundary Reynolds number Re$=4 u_{\mathrm{orb}} a_{\mathrm{orb}}/\nu_a$, where
$u_{\mathrm{orb}}$ and $a_{\mathrm{orb}}$ are the significant
surface orbital velocity and displacement amplitudes, and $\nu_a$
is the air viscosity, we take, for Re less than a critical value Re$_{c}$
\begin{equation}
S_{\mathrm{out}}\left(f,\theta\right) = - C_{\mathrm{dsv}}
\frac{\rho_a}{\rho_w}\left\{ 2 k \sqrt{2 \nu \sigma}\right\}
F\left(f,\theta\right)\label{Sds_turb},
\end{equation}
where the constant $C_{\mathrm{dsv}}$ is equal to 1 in \cite{Dore1978}'s laminar theory. 

When the boundary layer is expected to be turbulent, for Re$\geq$Re$_c$, we take
\begin{equation}
S_{\mathrm{out}}\left(f,\theta\right) = -\frac{\rho_a}{\rho_w}\left\{  16 f_e \sigma^2 u_{\mathrm{orb}} / g
\right\} F\left(f,\theta\right)\label{Sds_visc}. 
\end{equation}
A few tests have indicated that a threshold Re$_{c}=2 \times 10^5$~m$/H_s$ provides reasonable result, although it may be a also be a function of the wind speed, and we have no explanation for the dependence on $H_s$. A constant threshold close to $2 \times 10^5$ provides similar results. 
Here we shall use $C_{\mathrm{dsv}}=1.2$, but the results are not too sensitive to the exact value.

The parameterization of the turbulent boundary layer is more problematic. Without direct measurements in the boundary layer, there is ample room for speculations. From the analogy with an oscillatory boundary layer over a fixed bottom \citep{Jensen&al.1989}, the values of $f_e$ inferred from the swell observations, in the range 0.004 to 0.013 \citep{Ardhuin&al.2009}, correspond to a surface with a very small roughness. Because, we also expect the wind to influence $f_e$, the 
parameterization form includes adjustable effects of wind speed on the roughness, and an explicit correction of $f_e$. This  latter correction takes the form of a Taylor expansion to first order in $u_\star / u_{\mathrm{orb}}$, 
\begin{equation}
f_e = s_1\left\{ f_{e,GM} + \left[\left|s_3\right| 
+ s_2 \cos(\theta-\theta_u)\right]\frac{u_\star }{ u_{\mathrm{orb}}}\right\},\label{fevar}
\end{equation}
where $f_{e,GM}$ is the friction factor given by Grant and
Madsen's (1979)\nocite{Grant&Madsen1979} theory for rough
oscillatory boundary layers without a mean flow. 
Adequate swell dissipation is obtained with constant values of $f_e$ in the range 0.004 to 0.007, but these do not necessarily produce the best results when comparing wave heights to observations. Based on the simple idea that most of the air-sea momentum flux is supported by the pressure-slope correlations that give rise to the wave field \citep{Donelan1998,Peirson&Banner2003}, we have set the surface roughness to
\begin{equation}
 z_0' = r_{z0} z_0 \label{rz0}
\end{equation}
where $r_{z0}$ is here set to 0.04, of the roughness for the wind. We thus give the more generic equation (\ref{fevar}) for $f_e$, with $f_{e,GM}$ of the order of 0.003 for values of $a_{\mathrm{orb}}/z_0'$ of the order of 2$\times 10^5$.

\begin{figure}[htb]
\centerline{\includegraphics[width=0.7\linewidth]{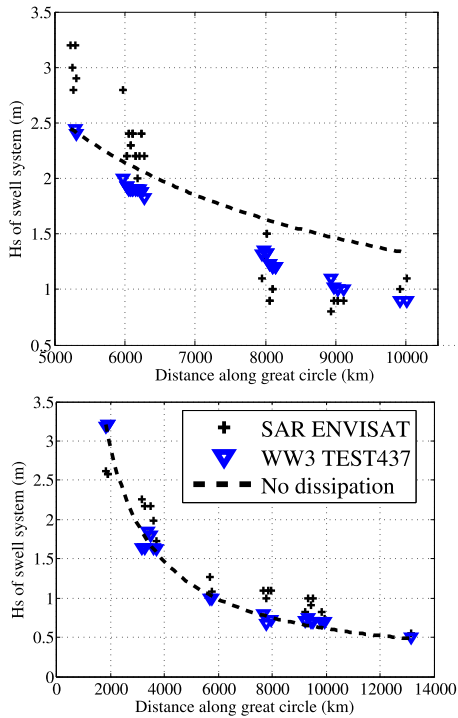}}
\caption{Comparison of modelled swell significant heights, following the propagation of the two swells shown by Ardhuin et al. (2009) with peak periods of 15~s and high and low dissipation rates. Known biases 
in the level 2 data have been corrected following Collard et al. (2009).\label{swelldiss}}
\end{figure}

The coefficients $s_2$ and $s_3$ of the $O(u_\star / u_{\mathrm{orb}})$ correction have been adjusted to -0.018 and 0.015, respectively, the former negative value giving a stronger dissipation for swells opposed to winds, when $\cos(\theta-\theta_u) < 0$. 
This gives a range of values of $f_e$ consistent with the observations, and reasonable hindcasts of swell decay (Fig. \ref{swelldiss}), with a small underestimation of dissipation for steep swells. An increase of $s_1$ from 0.8 to 1.1 produces negative biases on $H_s$ of the order of 30\% at all oceanic buoys (40\% for a partial wave height estimated from a spectrum restricted to periods around 15~s), so that the magnitude of the swell dissipation cannot be much larger than chosen here. 
Further discussion and validation of the swell dissipation is provided by the global scale hindcasts in Section 4.

\subsection{Wave breaking}
Observations show that waves break when the orbital velocity at their crest
$U_c$  comes close to the phase speed $C$, with a ratio $U_c/C > 0.8$ for
random waves \citep{Tulin&Landrini2001,Stansell&MacFarlane2002,Wu&Nepf2002}. It is nevertheless difficult  to parameterize the breaking of 
random waves, since the only available quantity here is the spectral density. This density can be related to the orbital velocity variance in a narrow frequency band. This question is addressed in detail by \cite{Filipot&al.2010}. Yet, a proper threshold has to be defined for this quantity, and the spectral rate of energy loss associated to breaking has to be defined. Also, breaking is intricately related to the complex non-linear evolution of the
waves \citep[e.g.][]{Banner&Peirson2007}. 

These difficulties will be ignored here. We shall parameterize the spectral dissipation rate directly from the wave spectrum, in a way similar to WZB. 
Essentially we distiguish between spontaneous and induced breaking, the latter being caused by large scale breakers overtaking shorter waves, and causing them to be dissipated.  
For the spontaneous breaking we parameterize the dissipation rate directly from the spectrum, without the intermediate step of estimating a breaking probability. 

We started from the simplest possible dissipation term formulated in terms
of the direction-integrated spectral saturation $B \left(f\right)$ given by eq. (\ref{defB}), 
with a realistic threshold $B_{0r}=1.2 \times 10^{-3}$ corresponding to the onset
of wave breaking \citep{Babanin&Young2005}. This saturation parameter corresponds exactly to the $\alpha$ parameter defined by \cite{Phillips1958}.  The value $B_{0}=8 \times 10^{-3}$, given by Phillips,  corresponds to a self-similar sea state in which waves of all scales have the same shape, limited by the breaking limit. 

This view of the sea state, however, ignores completely wave directionality. Early tests of parameterizations based on this definition of $B$ indicated that the 
spectra were too narrow \citep{Ardhuin&LeBoyer2006}. This effect could be due to many errors. Because \cite{Banner&al.2002} introduced a directional width in their saturation to explain some of the variability in observed breaking probabilities, we similarly modify the definition of $B$. Expecting also to have different dissipation rates in different directions, we define a saturation $B'$ that would correspond, in deep water, to a normalized velocity variance projected in one direction (in the case ${\mathrm{s_B}}=2$), with a further restriction of the integration of directions controlled by $\Delta_\theta$, 
\begin{equation}
B'\left(f,\theta\right)=
\int_{\theta-\Delta_\theta}^{\theta+\Delta_\theta} k^3 cos^{\mathrm{s_B}}\left(\theta-
\theta^{\prime}\right) F(f,\theta^{\prime}) \frac{C_g}{ 2 \pi } \mathrm d
\theta^{\prime} \label{defBofkprime},
\end{equation}
Here we shall always use $\Delta_\theta=80^\circ$. As a result, a sea state with two systems of same energy but opposite direction will typically produce much less dissipation than a sea state with all the energy radiated in the same direction. 

We finally define our dissipation term as the sum of the saturation-based term of
\cite{Ardhuin&al.2008d} and a cumulative breaking term
$S_{\mathrm{bk,cu}}$, 
\begin{eqnarray}
S_{\mathrm{oc}}(f,\theta)& =  \sigma
 \frac{C_{\mathrm{ds}}^{\mathrm{sat}}}{B^2_r} \left[ \delta_d
\max\left\{ B\left(f\right) -
B_r,0\right\}^2 \right.
\nonumber \\
  +& \left(1-\delta_d \right) \left. \max\left\{B'\left(f,\theta \right)- B_r
 ,0\right\}^2\right]F(f,\theta)  \nonumber \\
  +& S_{\mathrm{bk,cu}}(f,\theta) + S_{\mathrm{turb}}(f,\theta) \label{Sds_all}.
\end{eqnarray}
where
\begin{equation}
B\left(f \right)=\max\left\{B'(f,\theta), \theta \in [0,2
\pi[\right\} \label{defBof}.
\end{equation}

\begin{figure}[htb]
\centerline{\includegraphics[width=0.6\linewidth]{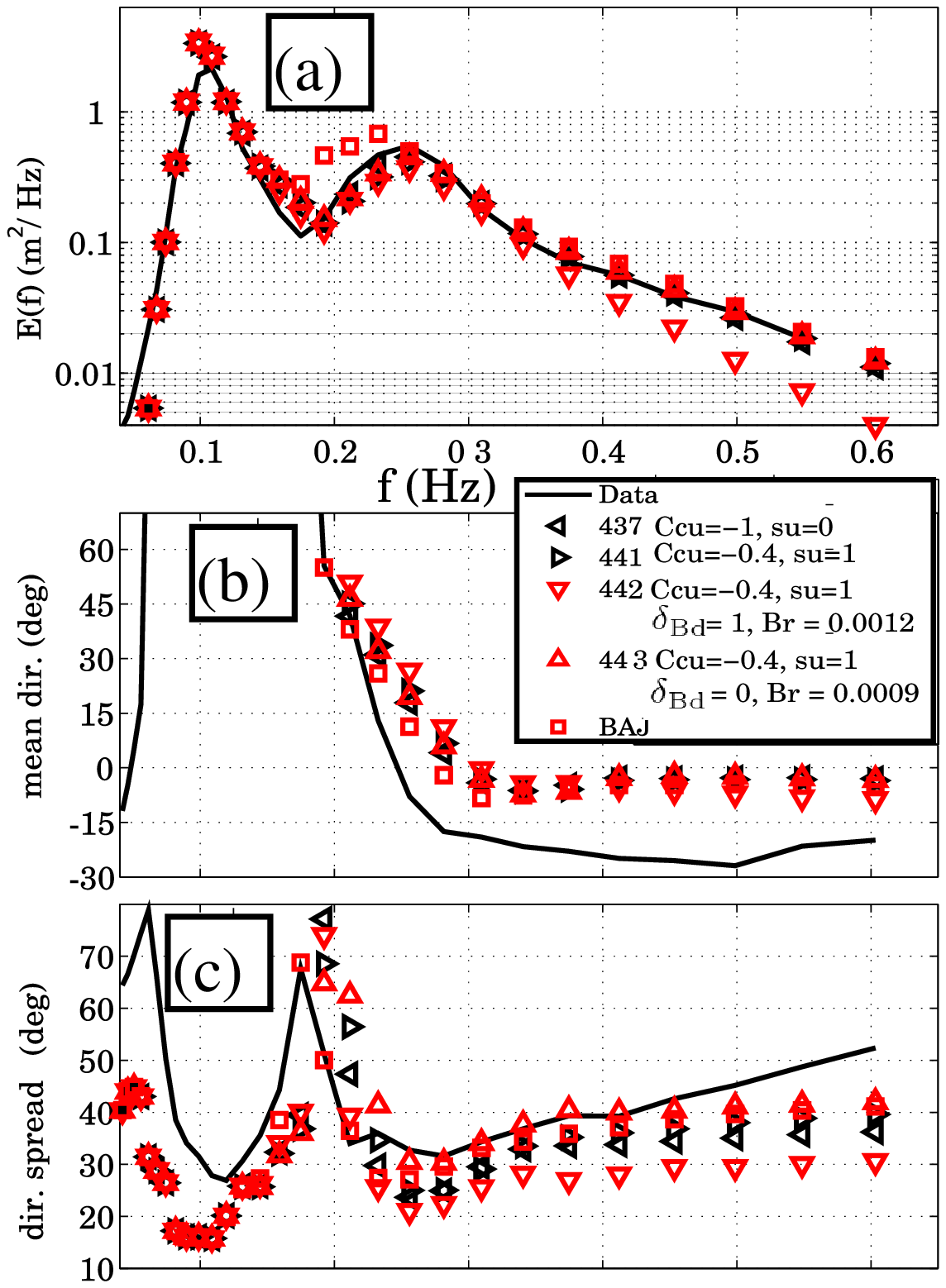}}
\caption{Wave spectra on 3 November 1999 at buoy X3 (fetch 39~km, wind speed $U_{10}=9.4$~m~s$^{-1}$), averaged over the time window 1200-1700
    EST, from observations and
model runs, with different model parameterizations (symbols): BAJ stands for \cite{Bidlot&al.2005}. (a) Energy, (b) mean direction (c) directional spread. This figure is analogue to the figures 10 and 11 in \cite{Ardhuin&al.2007}, the model forcing and setting are identical. It was further verified that halving the resolution from 1~km to 500~m does not affect the results. All parameters for BAJ, TEST441 and TEST443 are listed in tables 3 and 4. Input parameters for TEST443 are identical to those for TEST441, and TEST442 differs from TEST441 only in its isotropic direct breaking term, given by $s_B=0$, $\Delta_\theta=180^\circ$, and 
$B_r=0.0012$. It should be noted that the overall dissipation term in TEST443 is made anisotropic due to the cumulative effect, but this does not alter much the underestimation of directional spread.\label{fig:X3test}}
\end{figure}

\begin{figure}[htb]
\centerline{\includegraphics[width=\linewidth]{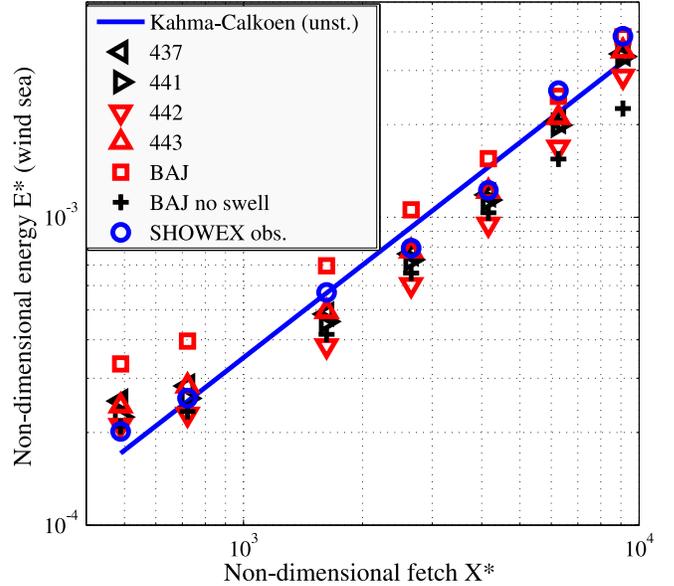}}
\caption{Fetch-limited growth of the windsea energy as a function of fetch on 3 November 1999, averaged over the time window 1200-1700 EST, from observations and
model runs, with different model parameterizations (symbols): BAJ stands for \cite{Bidlot&al.2005}. This figure is analogue to the figure 8 in \cite{Ardhuin&al.2007}, the model forcing and setting are identical. All parameters for BAJ, TEST441 and TEST443 are listed in tables A1 and A2. Input parameters for TEST443 are identical to those for TEST441, and TEST442 differs from TEST441 only in its isotropic direct breaking term, given by $s_B=0$, $\Delta_\theta=180^\circ$, and $B_r=0.0012$.\label{fig:Fetch}}
\end{figure}

The combination of an isotropic part (the term that multiplies $ \delta_d$) and a direction-dependent part (the term with  $1-\delta_d$) was intended to allow some control of the directional spread in resulting spectra. This aspect is illustrated  in figure \ref{fig:X3test} with a hindcast of the November 3 1999 case during the Shoaling Waves Experiment \citep{Ardhuin&al.2007}. Clearly, the isotropic saturation in the TEST442 dissipation (with the original threshold $B_r=0.0012$) produces very narrow spectra, even though it is known that the DIA parameterization for nonlinear interactions, used here, tends to broaden the spectra. The same behaviour is obtained with the isotropic parameterization by \cite{vanderWesthuysen&al.2007}, as demonstrated by \cite{Ardhuin&LeBoyer2006}. Further, using an isotropic dissipation at all frequencies yields an energy spectrum that decays faster towards high frequencies than the observed spectrum (Fig. \ref{fig:X3test}.a). On the contrary, a fully directional dissipation term (TEST443 with $\delta_d=0$) gives a better fit for all parameters. With $s_B=2$, we reduce  $B_r$ to  $0.0009$, a threshold for the onset of breaking
that is consistent with the observations of \cite{Banner&al.2000} and
\cite{Banner&al.2002}. 

The dissipation constant $C_{\mathrm{ds}}^{\mathrm{sat}}$ was adjusted to
$2.2\times 10^{-4}$ in order to give acceptable time-limited wave growth and reasonable directions in
fetch-limited growth \citep{Ardhuin&al.2007}. As noted in this previous work, similar growths of wave energy with fetch are possible with almost any magnitude of the wind input, but a reasonable mean direction in slanting fetch conditions selects the range of possible levels of input. Here the mean directions at the observed peak frequency  are still biassed by about 25$^\circ$ towards the alongshore direction with the parameterizations proposed here  (Fig. \ref{fig:X3test}.b),  which is still less that the 50$^\circ$ obtained with the weaker \cite{Tolman&Chalikov1996} source terms \citep[][figure 11]{Ardhuin&al.2007}. A relatively better fit is obtained with the BAJ parameterization. This is likely due to either the stronger wind input or the weaker dissipation at the peak. It is likely that both features of the BAJ parameterization are more realistic than what we propose here. 

Figure \ref{fig:Fetch} shows the fetch-limited growth in wave energy of various parameterizations. We repeat here the sensitivity test to the presence of swell, already displayed in \cite{Ardhuin&al.2007}. Whereas the 1~m swell causes an unrealistic doubling of the wind sea energy at short fetch in the BAJ parameterization, the new parameterizations, just like the one by \cite{vanderWesthuysen&al.2007} are, by design, insensitive to swell (not shown).

The dissipation $S_{\mathrm{turb}}$ due to wave-turbulence interactions  \citep{Ardhuin&Jenkins2006} is expected to be much weaker than all other terms and will be neglected here.

Finally, following the analysis by \cite{Filipot&al.2010}, the threshold $B_r$ is corrected for shallow water, so that $B'/B_r$ in different water depths corresponds to the same ratio of the root mean square orbital velocity and phase speed. 
For periodic and irrotational waves, the orbital velocity increases much more rapidly than the wave height as it approaches the breaking limit. Further, due to nonlinear distortions in the wave profile in shallow water, the height can be twice as large as the height of linear waves with the same energy. In order to express a relevant threshold from the elevation variance, we consider the slope $k H_{\mathrm{lin}}(kD)$ of an hypothetical linear wave that has the same energy as the wave of maximum height. In deep water\footnote{This value of the maximum equivalent linear height $H_{\mathrm{lin}}=2\sqrt{2E}$, with $E$ the elevation variance, is smaller than the usual value $k H=0.88$ due to the correction for the nonlinear wave profile for which $H > \sqrt{2E}$.}, $k H_{\mathrm{lin}}(\infty) \approx 0.77$, and for other water depths we thus correct $B_r$ by a factor $(kH_{\mathrm{lin}}(kD)/H_{\mathrm{lin}}(\infty))^2$. Using streamfunction theory \citep{Dalrymple1974}, a polynomial fit as a function of $Y=\tanh(kD)$ gives 
\begin{equation}
B'_r= B_r Y \left[M_4 Y^3   + M_3 Y^2 +  M_2 Y +    M_1 \right] \label{defBprime}.
\end{equation}
such that $B'_r=B_r$ in deep water. The fitted constants are $M_4=1.3286$, $M_3=-2.5709$ , $M_2=1.9995$ and 
$M_1=0.2428$. Although this behaviour is consistent with the variation of the depth-limited breaking parameter $\gamma$ derived empirically by \cite{Ruessink&al.2003}, the resulting dissipation rate is not yet expected to produce realistic results for surf zones because no effort was made to verify this aspect. This is the topic of ongoing work, outside of the scope of the present paper.  

The cumulative breaking term $S_{\mathrm{bk,cu}}$ represents the smoothing of the
surface by big breakers with celerity $C'$ that wipe out smaller
waves of phase speed $C$. Due to
uncertainties in the estimation of this effect in the observations of \cite{Young&Babanin2006b},
we use the theoretical model of \cite{Ardhuin&al.2009}. Briefly,
the relative velocity of the crests is the norm of the vector
difference, $\Delta_C =\left|\mathbf{C}-\mathbf{C}'\right|$, and
the dissipation rate of short wave is simply the rate of passage
of the large breaker over short waves, i.e. the integral of
$\Delta_C \Lambda(\mathbf{C}) d\mathbf{C}$, where $\Lambda
(\mathbf{C}) d\mathbf{C}$ is the length of breaking crests per
unit surface that have velocity components between $C_x$ and
$C_x+dC_x$, and between $C_y$ and $C_y+dC_y$ \citep{Phillips1985}.
Because there is no consensus on the form of $\Lambda$
\citep{Gemmrich&al.2008}, we prefer to link $\Lambda$ to breaking
probabilities. Based on  Banner et al. (2000, figure 6, $b_T=22 \left(\varepsilon-0.055\right)^2$), and
taking their saturation parameter $\varepsilon$ to be of the order
of $1.6 \sqrt{B'(f,\theta)}$, the breaking probability of dominant waves
waves is approximately
\begin{equation}
P=56.8\left(\max\{\sqrt{B'(f,\theta)}-\sqrt{B'_r},0\}\right)^2.\label{PBanner}
\end{equation}
However, because they used a zero-crossing analysis, 
for a given wave scale, there are many times when waves are not counted because the record is dominated by another scale: in their analysis there is only one wave at any given time.
This tends to overestimate the breaking probability by a factor of 1.5 to 2 \citep{Manasseh&al.2006}, compared to the present approach in which we consider that several waves (of different scales) may be present at the same place and time. 
We shall thus correct for this effect, simply dividing 
$P$ by 2.

With this approach we define the spectral density of crest length (breaking or not)  per unit surface
$l(\mathbf{k})$ such that $\int l(\mathbf{k}) \mathrm{d}k_x \mathrm{d}k_y$ is the total length of all crests per unit surface, with a crest being defined as a local maximum of the elevation in one horizontal direction. In the wavenumber vector spectral space we take 
\begin{equation}
l(\mathbf{k})= 1/(2\pi^2 k)
\end{equation}
 which is equivalent to a constant in wavenumber-direction space $l(k,\theta)= 1/(2\pi^2)$. This number was obtained by 
 considering an ocean surface full of unidirectional waves, with one crest for each wavelength $2 \pi/k$ for each spectral 
 interval $\Delta k=k$, e.g. one crest corresponding to spectral components in the range 0.5~$k$ to 1.5~$k$. We further double
 the potential number of crests to account for the directionality of the sea state. These two assumptions have not been verified and thus the resulting value of $l(\mathbf{k})$ is merely an adjustable order of magnitude. 

Thus the spectral density of breaking crest length per unit surface is 
$\Lambda(\mathbf{k})=l(\mathbf{k})P(\mathbf{k})$. Assuming that any breaking wave instantly dissipates all the energy of all waves with frequencies higher by a factor $r_{\mathrm{cu}}$ or more, then the cumulative dissipation rate is simply given by the rate at which these shorter waves are taken over by larger breaking waves, times the spectral density, namely 
\begin{equation}
S_{\mathrm{bk,cu}}(f,\theta) = C_{\mathrm{cu}}  F \left(f,\theta\right) \int_{f' < r_{\mathrm{cu}} f } \Delta_C \Lambda(\mathbf{k'}) \mathrm{d\mathbf{k'}},
\label{Sds_cu1}
\end{equation}
where $r_{\mathrm{cu}}$ defines the maximum ratio of the frequencies of long waves that will wipe out short waves.

We now obtain $\Lambda$ by extrapolating eq. (\ref{PBanner}) to higher frequencies, 
\begin{eqnarray}
S_{\mathrm{bk,cu}}(f,\theta) = &C_{\mathrm{cu}}  F \left(f,\theta\right)
\int_0^{ r_{\mathrm{cu}} f }\int_0^{2\pi}\frac{28.4}{\pi}
\nonumber \\
\times \max &\left\{\sqrt{B(f',\theta')}-\sqrt{B_r},0\right\}^2
\frac{\Delta_C}{C_g'}\mathrm{d}\theta' \mathrm{d}f',\nonumber\\
\label{Sds_sat_isotropic}
\end{eqnarray}
We shall 
take $r_{\mathrm{cu}}=0.5$, and $C_{\mathrm{cu}}$ is a tuning coefficient expected to be a negative number of order 1, which also corrects for errors in the estimation of $l$. 

This generic form of the source terms produces markedly different balances for both mature  and fully developped seas. 
\begin{figure}[htb]
\centerline{\includegraphics[width=0.55\linewidth]{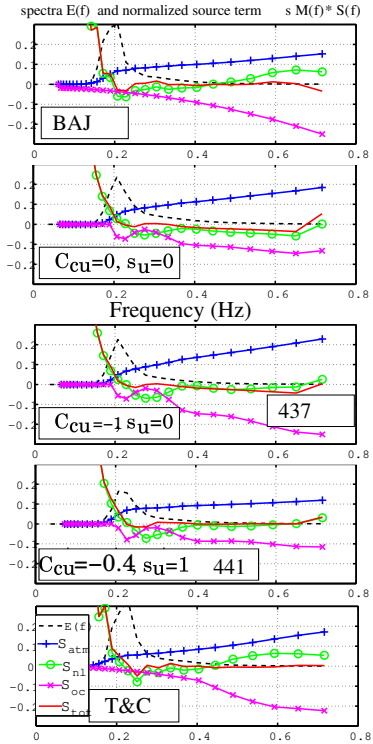}}
\caption{Academic test case over a uniform ocean with a uniform 10~m~s$^{-1}$ wind starting from rest, after 8 hours of integration, when $C_p/U_{10} \approx 1$. Source term balances given by the parameterization BAJ, and the parameterizations proposed here with the successive introduction of the cumulative breaking and the wind sheltering effects with the parameters $C_{\mathrm{cu}}$ and $s_u$. For BAJ, a  diagnostic $f^{-5}$ tail is applied above  2.5 the mean frequency. In order to make the high frequency balance visible, the source terms are multiplied by the normalization function $M(f)=\rho_w C /(\rho_a E(f) \sigma U_{10})$. The result with the parameterization of Tolman and Chalikov (1996) is also given for reference.\label{sourceterm_balance8h}}
\end{figure}
For mature seas, without cumulative effect, figure (\ref{sourceterm_balance8h}) shows that a balance is possible that gives roughly the same energy level and wind input term as the BAJ parameterization, up to 0.4~Hz. However, the balance for higher frquencies produces energy levels decreasing slower than $f^{-4}$ as the dissipation is too weak compared to the input, and thus the nonlinear energy flux is reversed, pumping energy from the tail to lower frequencies. 

The introduction of a strong cumulative term (TEST437) allows a balance at roughly the same energy level. However, with the present formulation this will lead to a dissipation too strong at high frequency for higher winds.  The introduction of the sheltering effect via the parameter $s_u$ (details in section \ref{wind_input_section}) is designed to get a balance with a weaker cumulative effect. 

The most important qualitative feature is the lack of a regular predefined shape for the normalized dissipation term $S_{oc}(f)/E(f)$. Whereas the shape given by $\delta k +(1-\delta) k^2$ is clearly visible in BAJ (with extremely high dissipation rates if one considers high frequencies), and the low to high frequency dissipation transition at 2 $f_p$ is evident in TC, the shape of the new dissipation rates are completely dictated by the local spectral saturation level. This leads to a relatively narrow peak of dissipation right above the spectral peak, where saturation is strongest. 

This feature helps to produce the realistic spectral shapes near the peak, with a steeper low frequency side and a more gentle slope on the high frequency side, contrary to the backward facing spectra produced by BAJ and TC. However, this localized strong relative dissipation, $S_{oc}(f)/E(f)$, is hard to reconcile with time and spatial scales of breaking events, and thus probably exaggerated. Indeed, there should be no significant difference in relative dissipation among the spectral components that contribute to a breaking wave crest, provided that they do not disperse significantly over the breaker life time, which is less than a wave period. Linear dispersion of waves with frequencies that differ by only 10\% should only produce small relative phase shift. Thus, there is no physical reason why a breaking event would take much more energy, relatively speaking, from the spectral band 1.1 to 1.2$f_p$ than from 1.2 to 1.3$f_p$. The factor 2 difference produced here in the relative dissipation rates is unrealistic. This strong relative dissipation at the peak (50\% higher than in BAJ) is one important factor that gives a slower growth of the wave spectrum in TEST441 compared to BAJ. It is possible that the localization of $S_{oc}$ at the peak compensates for the broader spectrum produced by the DIA compared to results with an exact non-linear interaction calculation.

\begin{figure}[htb]
\centerline{\includegraphics[width=0.65\linewidth]{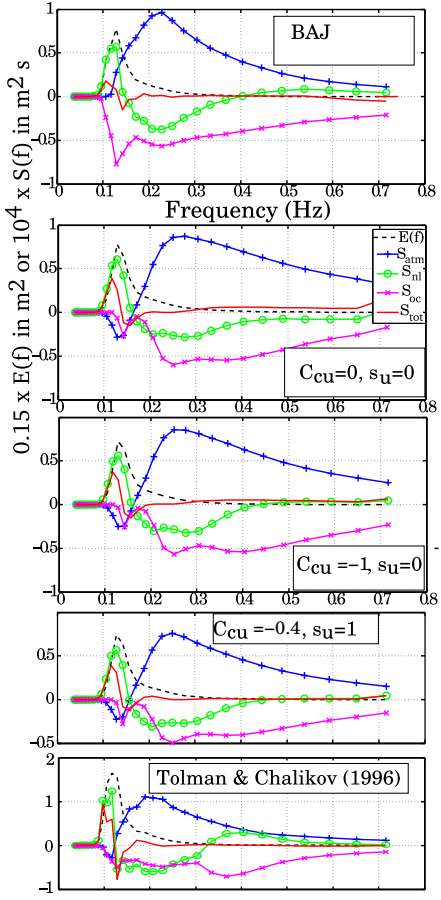}}
\caption{Same as figure (\ref{sourceterm_balance8h}) but after 48~h of integration and without normalization of the source terms.  Source term balances given by the parameterization BAJ, and the parameterizations proposed here with the successive introduction of the cumulative breaking and the wind sheltering effects with the parameters $C_{\mathrm{cu}}$ and $s_u$. For BAJ, a  diagnostic $f^{-5}$ tail is applied above  2.5 the mean frequency.\label{sourceterm_balance}}
\end{figure}
\begin{figure}[htb]
\centerline{\includegraphics[width=\linewidth]{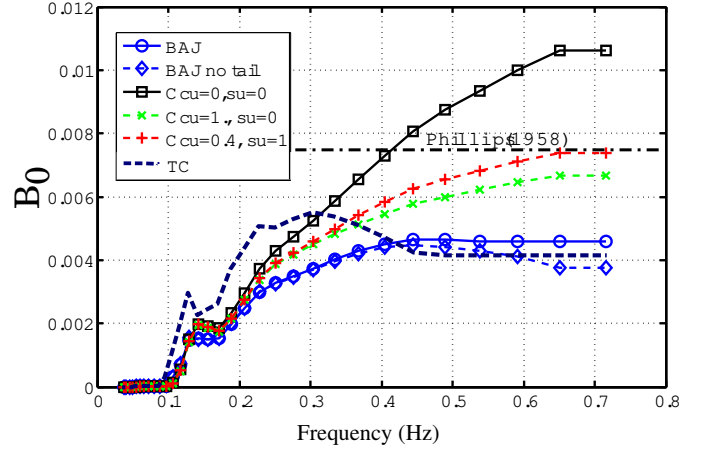}}
\caption{Values of the spectral saturation $B_0$ for the cases presented in figure \ref{sourceterm_balance}\label{saturations}.}
\end{figure}

We now consider ``fully developped'' conditions, illustrated by figure \ref{sourceterm_balance}, corresponding to long durations with steady wind and infinite fetch.
At low frequency, the non-linear swell damping term $S_{out}$ (the negative part of $S_{in}$) cancels about 30 to 50\% of the nonlinear energy flux, so that the sea state grows only very slowly. As a result ``full development'' does not exist (the wave height keeps growing), but the resulting energy is still compatible with the observations of mature sea states \citep{Alves&al.2003}. In contrast, the linear swell damping adjusted by \cite{Tolman2002d} to produce reasonable swell heights in the tropics is much smaller than the non-linear energy flux to low frequencies, even with the reduced interaction coefficient proposed by \cite{Tolman&Chalikov1996}. A non-linear swell dissipation appears necessary to obtain both a realistic damping of observed swells and a satisfatory agreement with mature wind waves. Nonlinearity also bring within the same order of magnitude 
the decay scales estimated for short \citep{Hogstrom&al.2009} and very long swells \citep{Ardhuin&al.2009}. 

Both parameterizations are physically very different from the parameterizations of the \cite{Komen&al.1984} family, including \cite{Bidlot&al.2005}. In these, the swell energy is lost to the ocean via whitecapping. Here we propose that this energy is lost to the atmosphere, with an associated momentum flux that drives the wave-driven wind observed in laboratories \citep{Harris1966} and for very weak winds at sea \citep{Smedman&al.2009}.  

  In the inertial range, a reasonable balance of all the source terms is obtained for $C_{\mathrm{cu}}=-0.4$ (figure 5). 
In this case, the spectrum approaches an $f^{-4}$ shape, for which $S_{nl}$ goes to zero, whereas in BAJ it decreases even 
faster than $f^{-5}$ which makes $S_{nl}$ positive above 0.4~Hz. The behaviour of the high frequency tail is best seen when displayed in non-dimensional form, as done in figure \ref{saturations}. This shows the unrealistic high level of the tail without cumulative effect nor modification of the wind input, and the equally unrealistic low tail with the 
BAJ parameterization, especially when the tail is left to evolve freely. 

Adding the cumulative effect can be used to control the tail level, but this degree of freedom is not enough. Indeed, in strongly forced conditions the dominant waves break frequently, and a high cumulative effect, $C_{\mathrm{cu}}=-1$, reduces the energy level in the tail below observed levels. This effect can be seen by considering satellite-derived mean square slopes (Fig. \ref{fig:mss}), or high moments of the frequency spectrum derived from buoy data (not shown but similar). 

That effect can be mitigated by decreasing $C_{\mathrm{cu}}$ or increasing $r_{\mathrm{cu}}$, so that dominant breaking waves will only wipe out much smaller waves. Instead, and because the wind to wave momentum flux was apparently too high in high winds, we chose to introduce one more degree of freedom, allowing a reduction of the wind input at high frequency.

\subsection{Wind input} \label{wind_input_section}
\begin{figure}[htb]
\centerline{\includegraphics[width=0.75\linewidth]{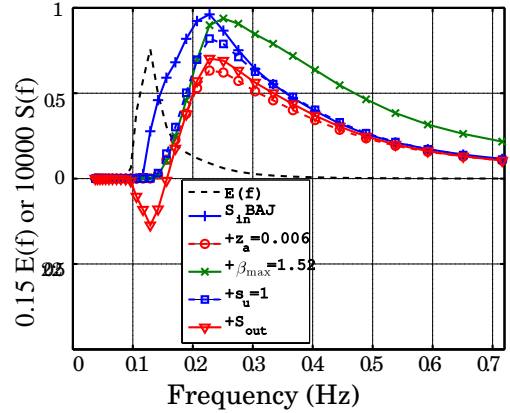}}
\caption{Incremental adjustements to the wind-wave interaction source term $S_{\mathrm{atm}}$, going from the BAJ form to the one used in 
TEST441. From one curve to the next, only one parameter is changed. The computations are performed for the same spectrum obtained by running the model from a calm sea for 8 hours with the BAJ parameterization and a wind speed of 10 m~s$^{-1}$. The reduction of $z_\alpha$ from 0.011 to 0.006 strongly reduces the input for frequencies in the range 0.15 to 0.2 Hz, which is probably overestimated in BAJ when average (20\%) levels of gustiness are considered: now the wind input goes to zero for $f=0.13$, which corresponds to $C/U_{10}=0.83$, whereas it is still significant at that wave age in the BAJ parameterization. As a result the much lower input level need a readjustment, performed here by increasing $\beta_{\max}$ to 1.52. Yet, this high value of $\beta_{\max}$ produces very high wind stress values and thus a very strong high frequency input. Adding the sheltering term $s_u=1$ allows a decent balance at high frequency. Finally, the addition of the air-sea friction term that gives swell dissipation produces a significant reduction of the input to the wind sea at $f=0.25$~Hz. It is questionable whether this mechanism also applies in the presence of the critical layer for those waves. This matter clearly requires more theoretical and experimental investigation.\label{Sin_new}}
\end{figure}

The wind input parameterization is thus adapted from \cite{Janssen1991} and the
following adjustments performed by \cite{Bidlot&al.2005,Bidlot&al.2007}.  The full wind input
source term reads
\begin{eqnarray}
S_{\mathrm{in}}\left(f,\theta\right) &=&
S_{\mathrm{in}}^{\mathrm{up}}(f,\theta) +
\frac{\rho_a}{\rho_w}\frac{\beta_{\mathrm{max}}}{\kappa^2}{\mathrm e}^{Z}Z^4
\left(\frac{u_\star}{C}\right)^2
\nonumber \\
& & \times \max \left \{ \cos(\theta - \theta_u),0\right \}^p \sigma F
\left(f,\theta\right),\label{SinWAM4}
\end{eqnarray}
where $\beta_{\mathrm{max}}$ is a non-dimensional growth parameter (constant),
$\kappa$ is von K\'{a}rm\'{a}n's constant. In the present implementation the
air/water density ratio is constant. The power of the cosine is taken constant
with $p=2$. We define the effective wave age $Z=\log(\mu)$ where $\mu$ is given by \cite{Janssen1991}, and corrected for intermediate water depths, so that
\begin{equation}
Z=\log(k z_1)+\kappa/\left[\cos\left(\theta - \theta_u\right) \left(u_\star/C +
z_\alpha \right)\right],\label{Z_Janssen}
\end{equation}
where $z_1$ is a roughness length modified by the wave-supported stress
$\tau_w$, and $z_\alpha$ is a wave age tuning parameter. $z_1$ is implicitly
defined by
\begin{eqnarray}
U_{10}&=&\frac{u_\star}{\kappa} \log\left(\frac{z_u}{z_1}\right) \label{U10}\\
z_0&=&\min\left\{\alpha_0 \frac{\tau}{g} , z_{0,\max}\right\} \label{z0}\\
z_1&=&\frac {z_0}{  \sqrt{1-\tau_w/\tau}},\label{z_1def}
\end{eqnarray}
where $z_u$ is the height at which the wind speed is specified, usually 10 meters. 
The maximum value of $z_0$ was added to reduce the unrealistic wind stresses at high
winds that are otherwise given by the standard parameterization. For example, 
$z_{0,\max}=0.0015$ is equivalent to setting a maximum wind drag
coefficient of $2.5\times 10^{-3}$. For the TEST441 parameterization, we have adjusted $z_\alpha=0.006$
and $\beta_{\max}=1.52$ (Fig. \ref{Sin_new}).

An important part of the parameterization is the calculation of the
wave-supported stress $\tau_w$, which includes the resolved part of the
spectrum, as well as the growth of an assumed $f^{-5}$ diagnostic tail beyond
the highest frequency. This parameterization is highly sensitive to the high
frequency part of the spectrum since a high energy level there will lead to a
larger value of $z_1$ and $u_\star$, which gives a positive feedback and reinforces the energy levels. 

In order to allow a balance with the saturation-based
dissipation, the wind input at high frequency is reduced by modifying 
the friction velocity $u_\star$. This correction also reduces the drag coefficient at high winds.
Essentially, the wind input is reduced for high frequencies and high winds, loosely following \cite{Chen&Belcher2000}.
This is performed by replacing $u_\star$ in eq. (\ref{SinWAM4}) with a frequency-dependent $u_\star'(f)$ defined by
\begin{eqnarray}
\left(u_\star '\right)^2&=&u_\star^2 \left(\cos \theta_u, \sin \theta_u
\right)  \nonumber \\
-& \left|s_u\right|&  \int_0^k \int_0^{2 \pi} \frac{S_{in}\left(f',\theta
\right)}{C} \left(\cos \theta, \sin \theta \right) {\mathrm d} f' \mathrm d
\theta,\nonumber
\\
\label{ustarp}
\end{eqnarray}

\noindent where the sheltering coefficient $\left|s_u\right|\sim 1$ can be used
to tune the wind stresses for high winds, which would be largely overestimated for
$s_u=0$. For $s_u > 0$ this sheltering is also applied within the diagnostic
tail, which requires the estimation of a 3-dimensional look-up table for the
high frequency stress. The shape of the new wind input is illustrated in figure \ref{Sin_new} for fully developped seas. 
Clearly, for relatively young waves the energy levels at the spectral peak are lower with $s_u=0.4$ (TEST441)  than in other runs, this is largely due to a reduced feedback of the wave age on the wind stress via the $\tau_w/\tau$ term in eq. (\ref{z_1def}).

\section{Consequences of the source term shape}
We have already illustrated the effects of various parameters on spectral shapes in academic time-limited and more realistic fetch-limited conditions. We now look at real sea states observed in the world ocean. 
Although wave spectra are difficult to compare to the few available observations, we have investigated the systematic variation of spectral moments 
\begin{equation}
m_n(f_c) = \int_0^{f_c} f^n E(f) {\mathrm d}f \label{mn}.
\end{equation}
with $n=2$,3 and 4, and cut-off frequencies in the range 0.2 to 0.4~Hz. Such moments are relevant to a variety of applications. \cite{Ardhuin&al.2009} investigated the third moment, which is proportional to the surface Stokes drift in deep water, and found that buoy data are very well represented by a simple function, which typically explains 95\% of the variance, 
\begin{eqnarray}
m_3(f_c)&\simeq& \frac{5.9 g U_{10}}{(2 \pi)^3} \times 10^{-4}
\left[1.25-0.25\left(\frac{0.5}{f_c}\right)^{1.3}\right] 
\nonumber \\
& \times& \min\left\{U_{10},14.5\right\} + 0.027\left(
H_s-0.4\right),\label{Ussnd_U10} \nonumber\\
\end{eqnarray}
where $f_c$ is in Hertz, $U_{10}$ is in meters per second, and
$H_s$ is in meters.

\begin{figure}[htb]
\centerline{\includegraphics[width=\linewidth]{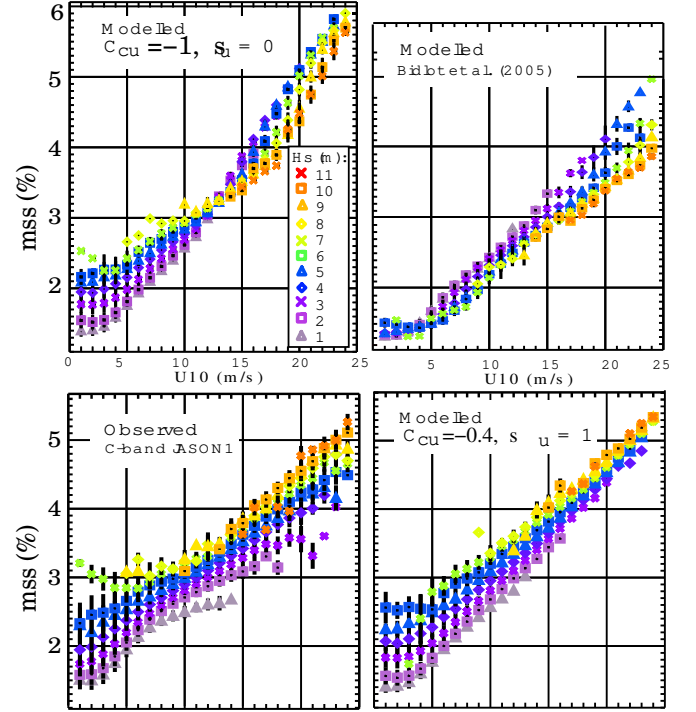}}
\caption{Variation of the surface mean square slope estimated as either $0.64/\sigma_0$ using the C-band altimeter on board JASON-1, after the correction of a 1.2 dB bias in the JASON data, or by integration of modelled spectra from 0 to 0.72~Hz, with either the $C_{\mathrm{cu}}=-0.4$ and $s_u=1$ parameterization (TEST441) or the parameterization by BAJ. For modelled values a constant 0.011 is added to account for the short waves that contribute to the satellite signal and that are not resolved in the model. This saturated high frequency tail is consistent with the observations of \cite{Vandemark&al.2004}. The original 1~Hz data from JASON is subsampled at 0.5~Hz and averaged over 10~s, namely 58~km along the satellite track. The same averaging is applied to the wave model result, giving the 393382 observations reported here, for the first half year of 2007.\label{fig:mss}}
\end{figure}

This relationship is well reproduced in hindcasts using $C_{\mathrm{cu}}=-0.4$ and $s_u=1$, while the BAJ source terms give a nearly  constant value of $m_3$ when $H_s$ varies and $U_{10}$ is fixed \citep{Ardhuin&al.2009}. 
Here we also consider the fourth moment $m_4$ which, for linear waves, is proportional to a surface mean square slope filtered at the frequency $f_c$. Figure \ref{fig:mss} shows that for any given wind speed mss$_C$ increases with the wave height \citep{Gourrion&al.2002}, whereas this is not the case of $m_4$ in the BAJ parameterization, or, for very high winds, when $C_{\mathrm{cu}}$ is too strong. In the case of BAJ, this is due to the $(k/k_r)^2$ part in the dissipation term (eq. \ref{WAM4dis}), which plays a role similar to the cumulative term in our formulation. For $C_{\mathrm{cu}}=-1$ and $s_u=0$, the cumulative effect gets too strong for wind speeds over 10~m~s$^{-1}$, in which case $m_4$ starts to decrease with increasing wave height, whereas for high winds and low (i.e. young) waves, the high frequency tail is too high and the mean square slope gets as large as 6\%, which is unrealistic. It thus appears, that the high frequency tail, for $s_u=0$, responds too much to the wind, hence our use of $s_u=1$ in the TEST441 combination. The presence of a cumulative dissipation term allows a different balance in the spectral regions above the peak, where an equilibrium range with a spectrum proportional to $f^{-4}$ develops \citep{Long&Resio2007}, and in the high frequency tail were the spectrum decays like $f^{-5}$ or possibly a little faster. The spectral level in the range 0.2 to 0.4~Hz was carefully compared against buoy data, and was found to be realistic. 

These interpretations of the model result assumes that the high frequency part of the spectrum can be simply converted to a wavenumber spectrum, using linear wave theory. This is not exactly the case as demonstrated by \cite{Banner&al.1989}. Also, there is no consensus on the nature of the spectrum modelled with the energy balance equation but, since non-resonant nonlinearities are not represented, the modelled spectra are expected to be more related to Lagrangian buoy measurements, rather than Eulerian measurements. This matter is left for further studies, together with a detailed interpretation of altimeter radar cross sections. Although it covers much less data, the analysis of $m_4$ obtained from buoy heave spectra produces results similar to figure \ref{fig:mss}.

\section{Verification}
In order to provide simplified measures of the difference between model time series $X_{\mathrm{mod}}$ and
observations $X_{\mathrm{obs}}$ we use the following definitions for the normalized root mean
square error (NRMSE), 
\begin{equation}
 \mathrm{NRMSE}(X)=\sqrt{\frac{\sum \left(X_{\mathrm{obs}}-X_{\mathrm{mod}}\right)^2}{\sum X_{\mathrm{obs}}^2}}
\end{equation}
the normalized bias, 
\begin{equation}
 \mathrm{NB}(X)=\sqrt{\frac{\sum X_{\mathrm{obs}}-X_{\mathrm{mod}}}{\sum X_{\mathrm{obs}}}}, 
\end{equation}
and Pearson's linear correlation  coefficient, 
\begin{equation}
 r(X)=\frac{\sum \left(X_{\mathrm{obs}}-\overline{X_{\mathrm{obs}}}\right) \left(X_{\mathrm{mod}}-\overline{X_{\mathrm{mod}}}\right) }{\sqrt{\sum \left(X_{\mathrm{obs}}-\overline{X_{\mathrm{obs}}}\right)^2 \left(X_{\mathrm{mod}}-\overline{X_{\mathrm{mod}}}\right)^2 }}, 
\end{equation}
where the overbar denotes the arithmetic average.

The normalisation of the errors allows a
quantitative comparison between widely different sea state regimes. Because
previous studies have often used (non-normalized) RMSE we also provide RMSE values.
In addition to the coastal fetch-limited case of SHOWEX, presented above, the parameterizations are calibrated on at 
the global scale and validated in two other cases. 

\subsection{Global scale results}
We present here results for the entire year 2007, using a stand-alone 0.5$^\circ$ resolution grid, covering the globe from 80$^\circ$ south to 80$^\circ$ north. The model has actually been adjusted to perform well over this data set, but the very large number of observations (over 2 million altimeter collocation points) makes the model robust, and an independent validation on 2008 gives identical results. The interested reader may also look at the monthly reports for the SHOM model \citep[e.g.][]{Bidlot2008}, generated as part of the model verification project of the IOC-WMO
Joint Commission on Oceanography and marine Meteorology (JCOMM), in which the TEST441 parameterization ($C_{\mathrm{cu}}=-0.4$ and $s_u=1$) is used, except for the Mediterranean where, the TEST405 has been preferred for its superior performance for younger seas. These SHOM models are ran in a combination of two-way nested grids  \citep{Tolman2007}. The monthly JCOMM reports include both analysis and forecasts, but, since they are produced in a routine setting, many SHOM calculations from December 2008 to June 2009 have been affected by wind file transfer problems.

\begin{figure*}
\begin{center}
\centerline{\includegraphics[width=\textwidth]{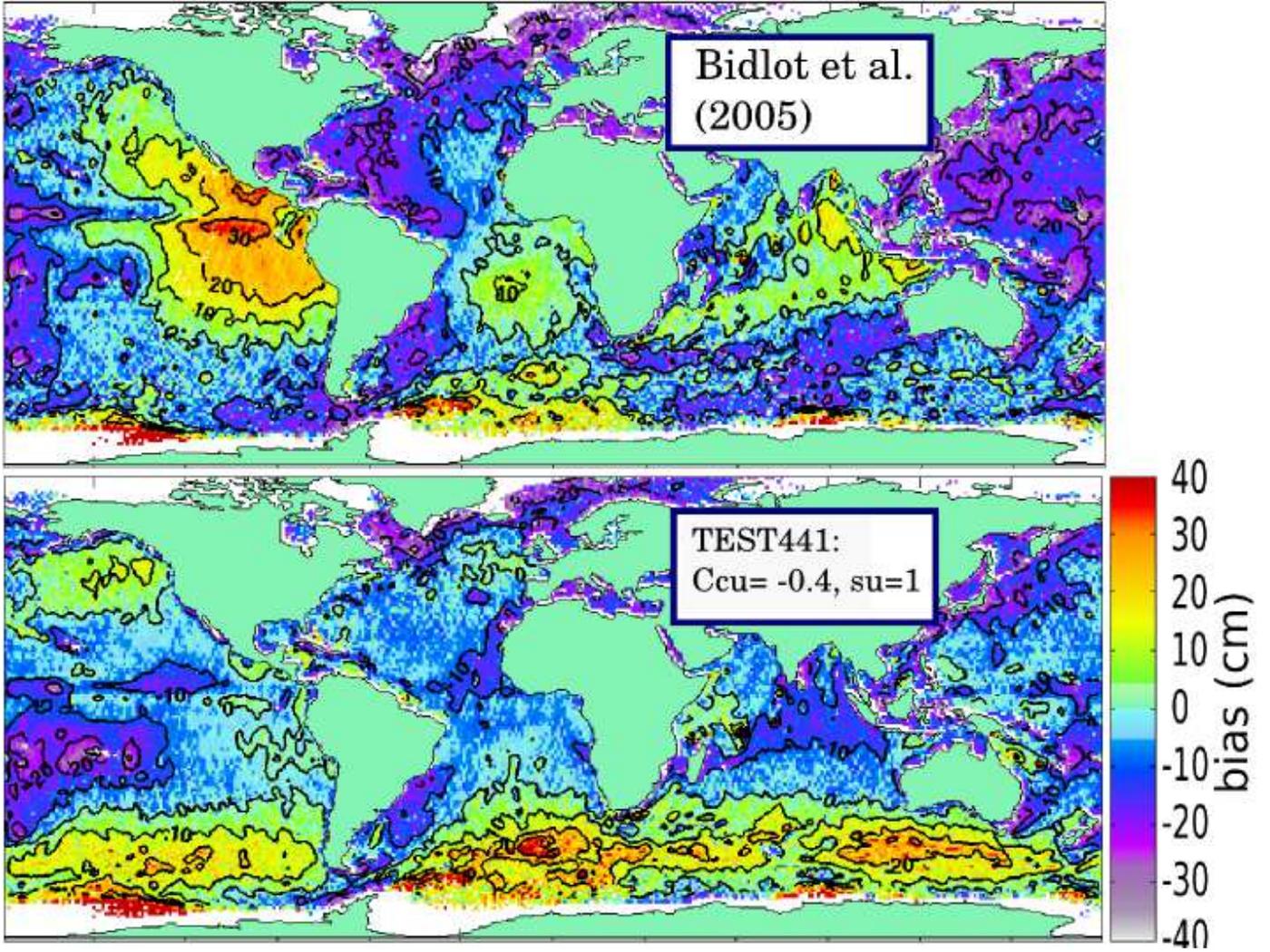}}
\caption{Bias for the year 2007 in centimeters. The global 0.5 WWATCH model is compared to altimeters JASON, ENVISAT and GFO following the
method of Rascle et al. (2008). The top panel is the result with the BAJ parameterization, and the bottom 
panel is the result with the $C_{\mathrm{cu}}=-0.4$ and $s_u=1$ (TEST441) parameterization.\label{Alti1}} 
\end{center}
\end{figure*}

\begin{figure}[htb]
\centerline{\includegraphics[width=0.5\linewidth]{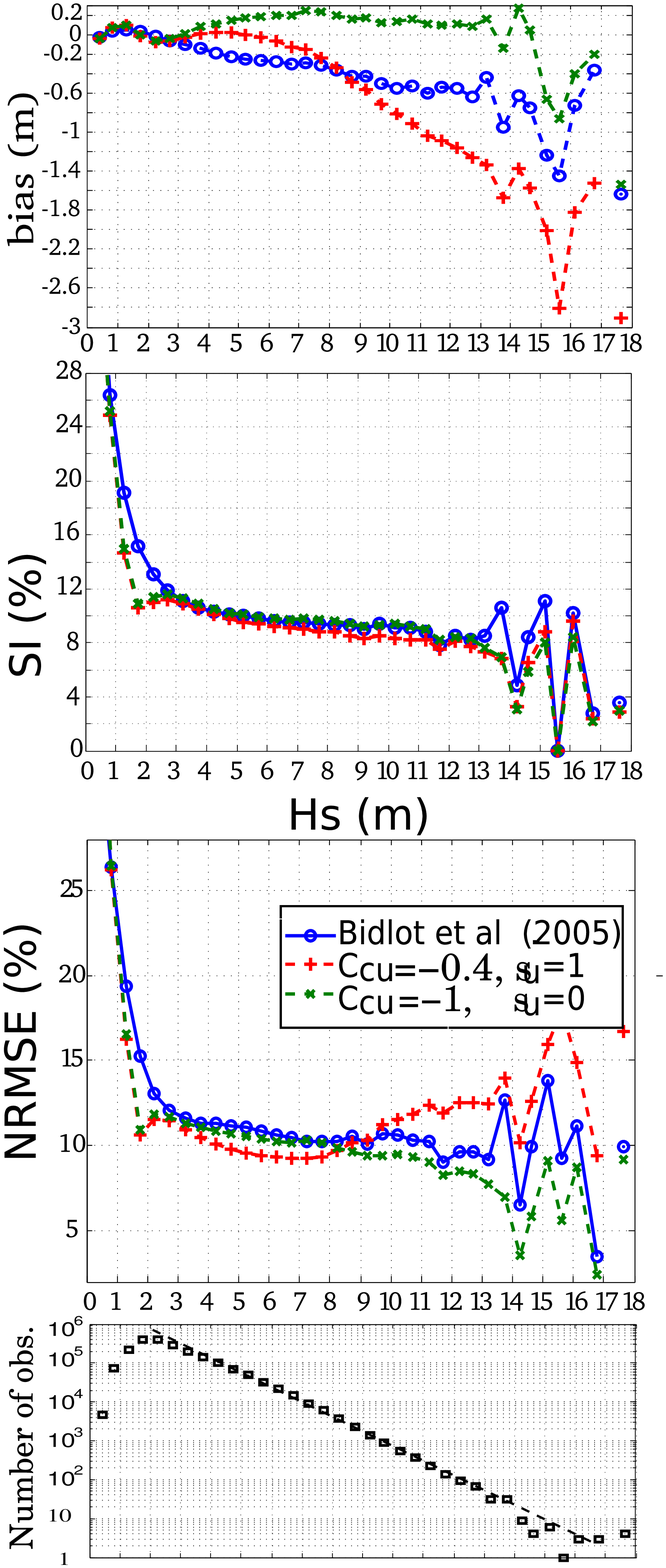}}
\caption{Wave model errors as a function of $H_s$. All model parameterizations are used in a global WWATCH model settings 
using a 0.5$^\circ$ resolution. The model output at 3h intervals is compared to JASON, ENVISAT and GFO following the
method of Rascle et al. (2008). Namely, the altimeter 1~Hz Ku band estimates of $H_s$ are averaged over 1$^\circ$. After this averaging, the total number of observations is 2044545. The altimeter estimates are not expected to be valid for $H_s$ larger than about 12~m, due to the low signal level and the fact that the waveform used to estimate $H_s$ is not long enough in these cases.\label{AltiHs}} 
\end{figure}

Comparing model results for $H_s$ to well-calibrated \citep{Queffeulou&CroizeFillon2008} altimeter-derived measurements provides a good verification of the model performance in a number of different wave climates. Figure (\ref{Alti1}) shows that, as expected, the important positive bias in the swell-dominated regions when using the BAJ parameterization, has been largely removed. This is essentially the signature of the specific swell dissipation that is parameterized in $S_{\mathrm{out}}$. The largest bias pattern now appears in the southern ocean, reaching 30~cm in the Southern Atlantic. Although this bias is small compared to the local averaged wave height, it is rather strange when the model errors are plotted as a function of wave height in figure (\ref{AltiHs}). Why would the model overestimate the Southern ocean waves but understimate the very large waves? 

The structure of the large bias, also seen in model results with BAJ, is reminiscent of the observed pattern in iceberg distribution observed by \cite{Tournadre&al.2008}. These observed iceberg distributions are enough to give a cross-section for incoming waves of the order of 1 to 10\% for a 250~km propagation length. Taking icebergs into account could actually reverse the sign of the bias. This matter will be investigated elsewhere. 

Also noticeable is a significant negative bias in the equatorial south Pacific, amplified from the same bias obtained with the BAJ parameterization. It is possible that the masking of subgrid islands \citep{Tolman2003} introduces a bias by neglecting shoreline reflections. This model defect could be exacerbated in this region by the very large ratio of shoreline 
length to sea area. This will also require further investigation. Finally, the negative biases for $H_s$ on mid-latitude 
east coasts are reduced but still persist. It is well known that these areas are also characterized by strong boundary 
currents (Gulf Stream, Kuroshio, Agulhas ...) with warm waters that is generally conducive to wave amplification and 
faster wind-wave growth \citep[e.g.][]{Vandemark&al.2001}. Neither effect is included in the present calculation because 
the accuracy of both modelled surface currents and air-sea stability parameterizations are likely to be insufficient \citep{Collard&al.2008,Ardhuin&al.2007}\nocite{Rascle&al.2008}. 

\begin{figure*}
\begin{center}
\centerline{\includegraphics[width=\linewidth]{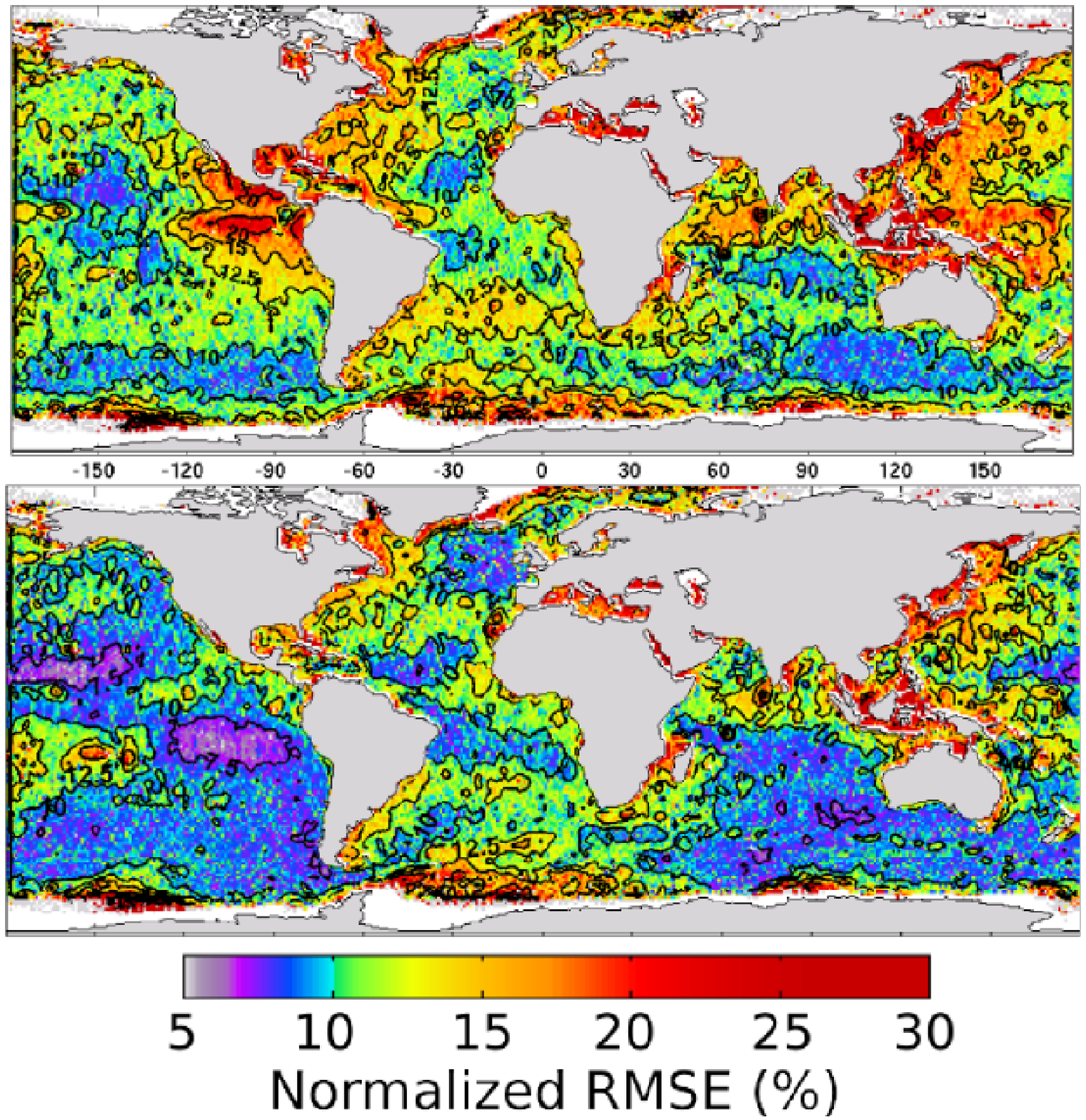}}
\caption{Normalized RMSE for the significant wave height over the year 2007, in percents. The global 0.5$^\circ$ resolution WWATCH model is compared to altimeters JASON, ENVISAT and GFO following the
method of Rascle et al. (2008). The top panel is the result with the BAJ parameterization, and the bottom 
panel is the result with the $C_{\mathrm{cu}}=-0.4$ and $s_u=1$ (TEST441) parameterization. \label{Alti2}}
\end{center}
\end{figure*}

The reduction of systematic biases clearly contributes to the reduction of r.m.s. errors, as evident in the equatorial east Pacific (Fig. \ref{Alti2}). However, the new parameterization also brings a considerable reduction of scatter, with reduced errors even where biases are minimal, such as the trade winds area south of Hawaii, where the NRMSE for $H_s$ can be as low as 5\%. When areas within 400~km from continents are excluded, because the global model resolution may be inadequate, significant errors ($> 12.5$\%, in yellow to red)  remain in the northern Indian ocean, on the North American and Asian east coasts, the Southern Atlantic. The parameterizations TEST405, TEST437 and TEST441 produce smaller errors on average than BAJ. It is likely that the model benefits from the absence of swell influences on wind seas: swell in BAJ typically leads to a reduced dissipation and stronger wind wave growth. As models are adjusted to average sea state conditions, this adjustement leads to a reduced wind sea growth on east coasts where there is generally less swell. 

Although much more sparse than the altimeter data, the in situ measurements collected and exchanged as part as the JCOMM wave model verification is very useful for constraining other aspects of the sea state. This is illustrated here with mean periods $T_{m02}$ for data provided by the U.K. and French meteorological services, and peak periods $T_p$ for all other sources. It is worth noting that the errors on $H_s$ for in situ platforms are comparable to the errors against altimeter data.

\begin{figure*}
\begin{center}
\centerline{\includegraphics[width=0.7\linewidth]{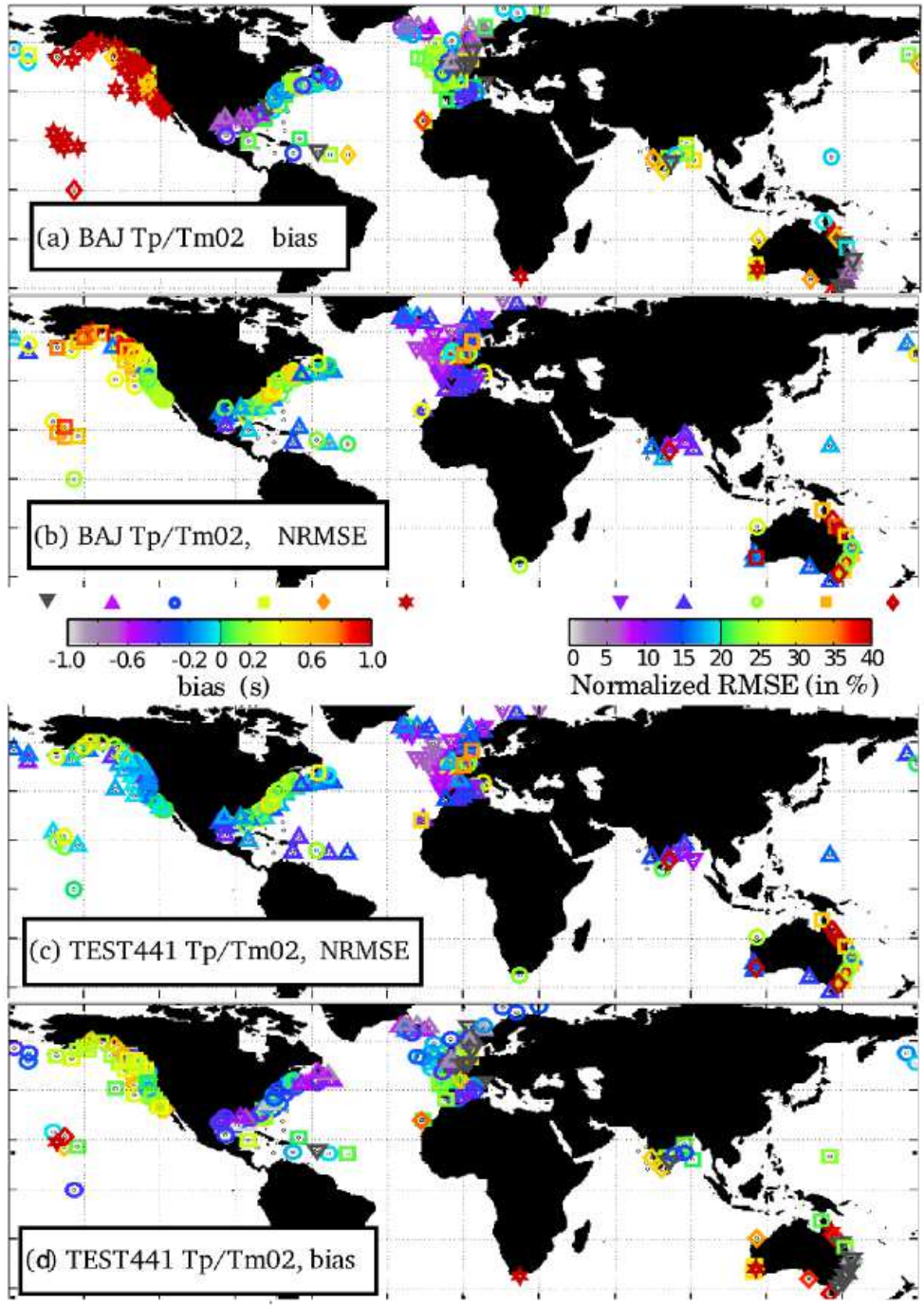}}
\caption{Statistics for the year 2007 based on the JCOMM verification data base \citep{Bidlot&al.2007b}. 
Bais (a,d) and NRMSE (b,c) for $T_p$ or
$T_{m02}$ at in situ locations  using the BAJ or the proposed TEST441 ($T_p$ is shown at all buoys except U.K. and French
buoys for which $T_{m02}$ is shown). 
The different symbols are only used to help distinguish the various colors and do not carry extra information.
\label{buoyvalid2}}
\end{center}
\end{figure*}

With the BAJ parameterizations, the largest errors in the model results are the large biases on peak periods on the U.S. West coast (Fig. \ref{buoyvalid2}), by 1.2 to 1.8~s for most locations, and the understimation of peak periods on the U.S. East Coast. However, peak and mean periods off the European coasts were generally very well predicted. When using the TEST441 parameterization, the explicit swell dissipation reduces the bias on periods on the U.S. West coast, but the problem is not completely solved, with residual biases of 0.2 to 0.4~s. This is consistent with the validation using satellite SAR data (Fig. 1), that showed a tendency to underpredict steep swells near the storms and overpredict them in the far field. A simple increase of the swell dissipation was tested but it tended to deteriorate the results on other parameters. 
On European coasts, despite a stronger bias, the errors on $T_{m02}$ are particularly reduced. Again, this reduction of the model scatter can be largely attributed to the decoupling of swell from windsea growth. 

 \begin{table}[htbp]
  \centering
  \caption[]{Model accuracy for measured wave parameters over the oceans in 2007. mss data from JASON 1 corresponds to January to July 2007 (393382 co-located points).
 Unless otherwise specified by the number in parenthesis, the cut-off frequency is take to be 0.4~Hz, $C$ stands for C-band. The normalized bias (NB) is defined as the bias divided by the r.m.s. observed value, while the scatter
 index (SI) is defined as the r.m.s. difference between modeled and observed values, after correction for the bias, normalized by the r.m.s. observed
 value, and $r$ is Pearson's correlation coefficient. These global averages are area-weighted, and the SI and NRMSE are the area-weighted averages of the local SI and NRMSE.\label{table_glob}}
  \begin{tabular}{lcccc}
\hline
\hline
               & BAJ     &  TEST  & TEST    & TEST  \\
               &         &       405   & 437    & 441 \\
\hline
$H_s$          & & & &  \\
NB(\%)         &  -2.1   &       -0.8  &  0.2 &  -1.23    \\
SI(\%)         &  11.8   &       10.5    &   10.6    &     10.4  \\
NRMSE(\%)      &   13.0  &      11.5     & 11.6        & 11.3      \\
\hline
$m_4(C)$       & & & & \\
NB(\%)         & -16.1   &       -4.9     &  -2.3 & -2.5  \\
SI(\%)         &  10.7   &       9.1     & 9.1    &  9.1   \\
 $r$           &  0.867  &       0.925   & 	0.931 & 0.939  \\
\hline
$T_p$          & & & & \\
NRMSE(\%)      &  24.1   &       19.0     & 19.4  &  18.2   \\
\hline
$T_{m02}$      & & & & \\
NRMSE(\%)      &  7.6    &       6.9     & 6.6  &  6.7   \\
 \hline
$m_3$          & & & & \\
NB(\%)         & -14.6   &       1.7     &  -2.3 & -2.4  \\
SI(\%)         &  20.6   &      12.6     &  14.8 & 12.6  \\
NRMSE(\%)      &  25.3   &      13.1     &  13.1 & 12.8  \\
$r$            &  0.934  &      0.971    & 0.961 & 0.973 \\
 \hline
 \hline
\end{tabular}
\end{table}

The general performance of the parameterizations is synthetized in table \ref{table_glob}. It is interesting to note that the  parameterization TEST405 that uses a diagnostic tail for 2.5 times the mean frequency gives good results in terms of scatter and bias even for parameters related to short waves ($m_3$, $m_4$). This use of diagnostic tail is thus a good pragmatic alternative to the more costly explicit resolution of shorter waves, which requires a smaller adapatative timestep, and more complex parameterizations. The diagnostic tail generally mimics the effect of both the cumulative and sheltering effects. Yet, the  parameterization TEST441 demonstrates that it is possible to obtain slightly better results with a free tail. The normalized biases indicated for the mean square slopes are only 
relative because of the approximate calibration of the radar cross section. They show that the BAJ parameterization \citep{Bidlot&al.2005}, and to a lesser extent the use of a $f^{-5}$ tail, produce energy levels that are relatively lower at high frequency. 

\subsection{Lake Michigan}
At the global scale, the sea state is never very young, and it is desirable to also verify the robustness of the parameterization in conditions that are more representative of the coastal ocean. We thus follow the analysis of wave model performances by \cite{Rogers&Wang2007}, hereinafter RW2007, and give results for the Lake Michigan, representative of relatively young waves. The model was applied with parameterizations BAJ, TC, TEST437 and TEST441 over the same time frame as investigated by RW2007, September 1 to November 14, 2002. The model setting and forcing fields are identical to the one defined by \citep{Rogers&al.2003}, with a 2~km resolution grid, a 10$^\circ$ directional resolution, and a wind field defined from in situ observations. The results at the position of National Data Buoys Center's buoy 45007 are compared to corresponding measurements. 

Using the directional validation method proposed by these authors, the TC parameterization understimates the directional spread $\sigma_\theta$ by 1.2 to 1.6$^\circ$ in the range 0.8 to 2.0 $f_p$, and more at higher frequencies. The understimation with BAJ is about half, and the TEST441 and TEST437 overpredict the directional spread by about 2.3 to 5.9$^\circ$ in the range 0.8 to 2.0$f_p$, and less so for higher frequencies. It thus appears that the broadening introduced to fit the SHOWEX 1999 observations is not optimal for other situations. A similar positive bias on directional spread is also found in global hindcasts. 

\begin{figure}[htb]
\centerline{\includegraphics[width=\linewidth]{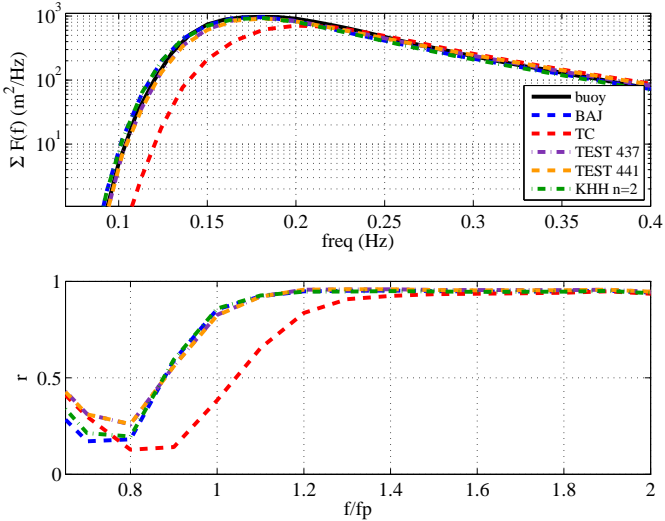}}
\caption{Model-data comparison at NDBC buoy 45007 (Lake Michigan). Top panel: Comparison of summed spectral density versus frequency for the duration of the simulation. Lower panel: Correlation coefficients versus normalized frequency (see text for explanation).\label{fig_LM}}
\end{figure}

Further results are presented in Table \ref{table_MICHIGAN} and Figure \ref{fig_LM}. The top panel of Figure \ref{fig_LM} compare the summed values of co-located model and observed spectral densities for the duration of the simulation. This presentation provides frequency distribution of bias of the various models, while also indicating the relative contribution of each frequency to the wave climate for this region and time period. The lower panel shows the correlation coefficient $r$ for the equivalent significant waveheights computed for multiple frequency bands. This is presented in terms of $f/f_p$ (bin width=0.1), with $f_p$ being calculated as the stabilized ``synthetic peak frequency'' of the corresponding buoy spectrum, as defined in RW2007. 

The most noticeable outcome of these comparisons is the relatively poor performance of the TC parameterizations. Taken in context with other TC results presented herein and prior undocumented application of the model in the Great Lakes with model wind fields, this suggests that these parameterizations have some undesirable dependence on scale, with the parameters adjusted by \cite{Tolman2002d} being most optimal for ocean-scale applications. 

The KHH parameterization perform well in this simple wave climate, consistent with prior published applications with the SWAN model, \citep{Rogers&al.2003} and RW2007, without the difficulties discussed above in mixed sea-swell conditions.

The BAJ, TEST437 and TEST441 models also perform well here. Taken together with the global comparisons above, we observe no apparent dependency of model skill on the scale of the application with these three physics. In the bias comparison, Fig. \ref{fig_LM} top panel, the BAJ model is nearly identical to the KHH model. Similarly, the two new models are also very close. Although TEST441 and TEST437 produce a minor underestimation of $H_s$ (Table \ref{table_MICHIGAN}), they give slightly better correlations with  observed wave heights and mean periods.

 \begin{table}[htbp]
  \centering
    \caption[]{Model-data comparison at NDBC buoy 45007 (Lake Michigan) for four frequency bands. Statistics are given for equivalent significant wave heights 
in the bands $0.5f_p  <f <0.8 f_p$ (band 1),  $0.8f_p  <f <1.2 f_p$ (band 2),  $1.2f_p  <f <2 f_p$ (band 3)  $2f_p  <f <3 f_p$ (band 4). The KHH $n=2$ run corresponds to the dissipation parameterization defined by \citep{Rogers&al.2003} based on \citep{Komen&al.1984} and applied in the WW3 code. For the peak frequency band, statistics are also given for the directional spread $\sigma_\theta$.\label{table_MICHIGAN} }
  \begin{tabular}{lccccc}
\hline
\hline
                      & BAJ           & TC       & TEST      & TEST  & KHH   \\
                      &               &          & 437       & 441   & $n=2$  \\
\hline
$H_s$ band 1           &               &          &           &       &       \\
SI(\%)                & 76            & 71       & 77        & 82    & 76    \\
$H_s$ band 2          &               &          &           &       &       \\
SI(\%)                & 16            &  19     & 15         & 15    & 15     \\
$H_s$ band 3           &              &         &             &       &        \\
SI(\%)                 & 18          & 26       & 17         & 17    & 19      \\
$H_s$ band 4           &              &          &             &      &         \\
SI(\%)                 & 20         & 32         & 18       & 17     & 22       \\
 \hline
$\sigma_\theta$ band 2  &               &          &           &       &       \\
SI(\%)                & 22            &  24     & 30         & 30    & 25     \\
bias ($^\circ$)         & -0.4         &  -1.6    &   2.3      & 2.6   & 0.8    \\
 \hline
$H_s$       
bias (m)               & -0.03       & -0.11    &   -0.02   & -0.02  & 0.04   \\
r                      &  0.95       &   0.94   &   0.96    &0.96    & 0.96    \\
SI(\%)                 &  19         &     25   &   18      & 18     & 19       \\
 \hline
$T_{m02}$               &              &          &           &        &      \\
bias (s)              &   0.02       &   -.39   &   -0.05   &  -0.05 & 0.01  \\
r                     &   0.89       &   0.87   &   0.90    & 0.90   & 0.90   \\
SI(\%)                &   10         &   14     &    9      & 9      & 10      \\
\hline
\hline
\end{tabular}
\end{table}

It thus appears that for such young seas, the directional spreading of the parameterization could be improved, but the energy content of various frequency bands, and as a result the mean period, are reproduced with less scatter than with previous parameterizations.

\subsection{Hurricane Ivan}
Although the global hindcast does contain quite a few extreme events, with significant wave heights up to 17~m, these were obtained with a relatively coarse wave model grid and wind forcing ($0.5^\circ$ resolution and 6-h timestep) that is insufficient to resolve small storms such as tropical cyclones \citep{Tolman&Alves2005}. Hurricane waves do share many similarities with more usual sea states \citep{Young2006}, but the high winds and their rapid rotation are particularly challenging for numerical wave models. 
It is thus necessary to verify that the new source functions perform adequately under extreme wind conditions. A simulation of Hurricane Ivan (Gulf of Mexico, September 2004) is chosen for this purpose because it was extensively measured \citep[e.g.][]{Wang&al.2005} and hindcasted.

Winds for this simulation are based on gridded surface wind analyses created by NOAA's Hurricane Research Division (HRD). These analyses are at three hour intervals, which for a small, fast-moving weather system is temporally too coarse to provide directly to the wave model. Therefore, as an intermediate step, fields are reprocessed to 30 minute intervals, with the storm position updated at each interval (thus, semi-Lagrangian interpolation). The wind speeds are reduced by factor 1/1.11 to convert from maximum sustained gust to hourly mean. The HRD winds do not cover the entire computational domain. For areas falling outside the domain, the nearest NDBC buoy wind observation is used. This produces some non-physical spatial discontinuities in the wind field, but these are smoothed in the wave model integration, and in any event, only affect weaker wind seas far from the 
storm center.

\begin{figure}[htb]
\centerline{\includegraphics[width=\linewidth]{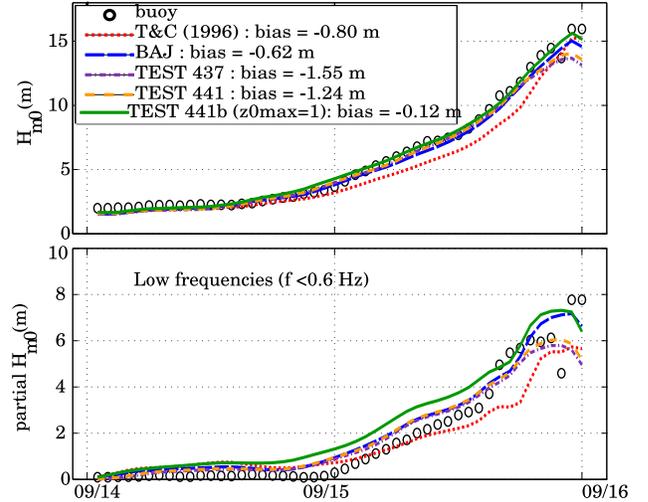}}
\caption{Time series of model and buoy significant wave height (top) and partial wave height (bottom, for $f< 0.06$ Hz), during the passage of Hurrican Ivan, at buoy 42040. The 10-m discus buoy was capized by waves and could not record waves after 16 September.\label{fig_Ivan}}
\end{figure}

Bathymetry is taken from the Naval Research Laboratory's 2 minute resolution database, DBDB2, coarsened to the model grid resolution (0.1 deg). The directional resolution is 10$^\circ$, and the frequency range is limited to 0.0418-0.4117~Hz. The model was applied from September 13 to September 16 2004.  Model results are illustrated by figure \ref{fig_Ivan}. The models were validated at all the buoys in the gulf of Mexico. Results from buoy 42040, where waves where largest, are shown here. 

Model runs with parameterizations BAJ, TEST437 and TEST441 give very similar results: close to the observations, except for the highest waves ($H_s >$  13 m at buoy 42040) where TEST437 and TEST441 give slightly smaller values. Results with the TC parameterization are generally lower in terms of $H_s$ than all these parameterizations that share a Janssen-type input. 
It appears that the new source term perform similarly to BAJ and are able to reproduce such young waves and severe sea states. 

Because the wind forcing enters the wave model through a wind stress parameterized in a way that may not apply to such
 conditions, it is worthwhile to re-examine some choices made above. In particular the surface roughness was allowed to exceed
 0.002 in TEST441b, which resulted in better estimates of $H_s$. Lifting this constraint shows that, for these very high 
winds,  the wind sheltering
 effect plays a similar role to the limitation of the roughness to $z_{0,\max}$, with the difference that it tends to
 narrow the wind input spectrum (Fig. \ref{Sin_new}). This narrower wind input has a limited effect of the wind stress 
and $H_s$, but is has a noticeable effect on the spectral shape. This is illustrated by the low
 frequency energy that appears to be strongly overestimated before the peak of the storm for TEST441b. In general the 
new parameterizations  provide results that are as reasonable as those of previous parameterizations, given the uncertainty of the wind forcing.

\section{Conclusions}
A set of parameterization for the dissipation source terms of the wave energy balance equation have been proposed, based on 
known properties of swell dissipation and wave breaking statistics. This dissipation includes an explicit nonlinear swell
 dissipation and a wave breaking parameterization that contains a cumulative term, representing the dissipation of short waves by
 longer breakers, and different dissipation rates for different directions. These dissipation parameterizations have been combined
 with a modified form of the wind input proposed by \cite{Janssen1991}, in which the questionable gustiness parameter $z_\alpha$
 has been reduced, and the general shape of the wind input has been significantly modified. The resulting source term balance is
 thus markedly different from the previous proposed forms, with a near-balance for very old seas between the air-sea friction
 term, that dissipates swell, and the nonlinear energy flux to low frequencies. Also, the wind input is concentrated in a narrower
 range of frequencies. 

For younger seas the wind input is relatively weaker than given by \cite{Janssen1991} but stronger than given by \cite{Tolman&Chalikov1996} (TC). However, the dissipation at the peak is generally stronger because it is essentially based on a local steepness and these dominant waves are the steepest in the sea state. As a result the short fetch growth is relatively weaker than with the source term combination proposed by \cite{Bidlot&al.2007} (BAJ). 
The choice of parameters tested here tend to produce broader directional spectra than observed in the Lake Michigan and global hindcasts, and slanting fetch directions that are too oblique relative to the wind (Fig. \ref{fig:X3test}). In this respect the new source terms are intermediate between BAJ and TC. 

Another likely defect comes from the definition of the saturation level used to define the breaking-induced dissipation. Here, as in the work by \cite{vanderWesthuysen&al.2007}, the saturation is local in frequency space, whereas wave breaking is naturally expected to have a relatively broad impact due to its localization in space and time \citep{Hasselmann1974}. This is expected to produce an overestimation of the energy just below the peak, and an understimation at the peak of the saturation spectrum. These effects likely contribute to the persistent overestimation of low frequency energy in the model. 

In spite of these defects, the new parameterization produces robust results and clearly outperform the \cite{Bidlot&al.2007} parameterization in global hindcasts, whether one considers dominant wave parameters, $H_s$, $T_{m02}$ and $T_p$ or parameters sensitive to the high frequency content, such as the surface Stokes $U_{ss}$ drift or the mean square slope. At global scales, errors on $H_s$, $T_p$ and $U_{ss}$ are - on average - reduced by 15, 25 and 50\% relative to those obtained with the parameterization by \cite{Bidlot&al.2007}. All global hindcasts results, for the year 2007 at least, are available for further analysis at 
ftp://ftp.ifremer.fr/ifremer/cersat/products/gridded/wavewatch3/HINDCAST/ .

Another important aspect was the validation at regional to global scales. We note that the TC paper does include verification with steady-state, fetch-limited growth curves. Though such verification is a useful step, the outcome of the Lake Michigan hindcast suggests that such verification does in no way anticipate skill in real sub-regional-scale applications. One of the parameterizations proposed here (TEST441) also gives slightly lower performance for young seas, which is not obvious in the case of lake Michigan, but was revealed by hindcasts of Mediterranean waves (not shown).

Because our intention was only to demonstrate the capability of new dissipation parameterizations and the resulting source term balances, we have not fully adjusted the 18 parameters that define the deep water parameterizations, compared to about 9 with  \cite{Bidlot&al.2007}. The results presented here are thus preliminary in terms of model performance, which is why the parameterizations are still given temporary names like TEST441. As illustrated by the hurrican Ivan hindcast, some parameters, such as $z_{0,\max}$, is probably unnecessary: in that particular case the removal of $z_{0,\max}$ improved the results, but for global scale results it had no impact at all (not shown). 

Because 5 of the extra parameters define the air-sea friction term that produces swell dissipation, and 2 define the cumulative breaking term, it is feasible to define a systematic adjustement procedure that should produce further improvements by separately adjusting swell, wind sea peak, and high frequency properties. In particular the directional distribution may be improved by  making the dissipation term more isotropic (i.e. taking $\delta_d > 0.3$) or modifying the definition of the saturation parameter $B'$ in equation (\ref{defBofkprime}). In part II we shall further investigate the response of the wave field to varying currents, from global scales to regional tidal currents. It is particularly expected that wave steepening will produce much more dissipation due to breaking, as envisaged by \cite{Phillips1984}.

Obviously, it is well known that the Discrete Interaction Approximation used here to compute the non-linear interactions is the source of large errors, and further calculations, will be performed using a more accurate estimation of these interactions in part III. 
\begin{acknowledgment}
 This research would not have been possible without the dedication of Hendrik Tolman, Henrique Alves, and Arun Chawla in putting together the core of the {WAVEWATCH-III$^{\rm{TM}}$} code. Wind and wave data were kindly
provided by ECMWF, M{\'e}t{\'e}o-France, and the French Centre
d'Etudes Techniques Maritimes Et Fluviales (CETMEF). The SHOM buoy deployments were managed by David Corman
with precious help from Guy Amis. Wave data were kindly provided by the U.S. Navy, ESA and CNES, and the many \textit{in situ} contributors to the JCOMM (WMO-IOC) exchange program coordinated by Jean-Raymond Bidlot. 
\end{acknowledgment}

\ifthenelse{\boolean{dc}}
{}
{\clearpage}
\begin{appendix}[A]
\section*{\begin{center}Parameter settings\end{center}}
\begin{table*}[h] \begin{center}
\begin{tabular}{lccccccc} \hline \hline
Parameter   & see eq. & variable in code  & WAM-Cycle4 & BAJ  & TEST405 & TEST437 & TEST441 \\
\hline
  $\alpha_0$ &  (\ref{SinWAM4}) & ALPHA0               & 0.01    & 0.0095 & idem &  idem &  idem \\
  $\beta_{\mathrm{max}}$ & (\ref{Z_Janssen}) & BETAMAX   & 1.2     & 1.2    & 1.55   &  1.52   &  idem \\
  $z_\alpha$ & (\ref{Z_Janssen}) & ZALP                 & 0.011  & 0.011 & 0.006  &  idem  &  idem \\
  $z_{0,\max}$ & (\ref{z0}) & Z0MAX               & N.A.   & N.A. & 0.002  &  idem & idem \\
  $s_u$ & (\ref{ustarp}) &TAUWSHELTER               & 0.0     & 0.0    & 0.0    &  0.0    &  1.0   \\
  $s_0$ &  &SWELLFPAR                 & 0       & 0      & 3      &  idem      & idem   \\
  $s_1$ & (\ref{fevar}) & SWELLF                    & 0.0     & 0.0    & 0.8    & idem    &  idem \\
  $s_2$ & (\ref{fevar})  & SWELLF2                   & 0.0     & 0.0    & -0.018  &  idem  &  idem \\
  $s_3$ & (\ref{fevar}) & SWELLF3                   & 0.0     & 0.0    & 0.015   &  idem   &  idem \\
  $0.5 \mathrm{Re}_c H_s $&  &SWELLF4          & 0.0     & $1 \times 10^5$     & idem    &  idem &  idem \\
  $C_{\mathrm{dsv}}$ & (\ref{Sds_turb}) &SWELLF5      & 0.0     & 0.0    & 1.2    &  idem  &  idem \\
  $r_{z0}$ & (\ref{rz0}) &Z0RAT                     & 0.0     & 0.0    & 0.04   &  idem  &  idem \\
  $z_u$    & (\ref{U10}) & ZWND                     & 10     & 10    & 10   &  10  &  10  \\
  $p$      & (\ref{SinWAM4}) & SINTHP                    & 2     & 2    & 2   &  2  &  2  \\
\hline
\end{tabular} \end{center}
\caption{Wind-wave interaction parameters as implemented in version 3.14-SHOM of the
WAVEWATCH III code, and values used in the tests presented here. In WWATCH, all parameters are accessible via the SIN3 namelist. All of these parameters are included in version 3.14 of WWATCH. $s_0$ is a switch that, if nonzero, activates the calculation of  $S_{\mathrm{out}}$.}
\label{tab:WAM4_parSIN}
\end{table*}

\begin{table*} \begin{center}
\begin{tabular}{lcccccccc} \hline \hline
Parameter      & see eq. &  variable in code                 & WAM4 & BAJ  & TEST405 & TEST437 & TEST441 & TEST443 \\
\hline
  $C_{\mathrm{ds}}$            & (\ref{WAM4dis})    &  SDSC1          & -4.5 & -2.1&  0.0 &  idem &  idem & idem \\
  $r$                          & (\ref{WAM4dis})    &  WNMEANP        & -0.5 & 0.5 &  0.5 &  idem &  idem & idem \\
  $f_{\mathrm{FM}}$            & (\ref{def_fc})    &  FXFM3          & 2.5  & 2.5 &  2.5 &  9.9 &  idem & idem \\
 $\delta_1$                    & (\ref{WAM4dis})  &  SDSDELT       & 0.5  & 0.4 &  0.0 & idem &  idem & idem \\
  $\delta_2$                   & (\ref{WAM4dis})    &  SDSDELTA2      & 0.5  & 0.6 &  0.0 & idem &  idem & idem \\
  $C_{\mathrm{ds}}^{\mathrm{sat}}$& (\ref{Sds_all}) & SDSC2           & 0.0  & 0.0 &$-2.2\times 10^{-5}$ & idem & idem & idem \\
  $C_{\mathrm{lf}}$            &     &  SDSLF          & 1.0  & 1.0 &  0.0& idem &  idem & idem \\
  $C_{\mathrm{hf}}$            &     &  SDSHF          & 1.0  & 1.0 &  0.0& idem &  idem & idem \\
  $\Delta_\theta$              &  (\ref{defBofkprime})   &  SDSDTH         & 0.0  & 0.0 &  80 & idem &  idem & idem \\
  $B_r$                        &   (\ref{defBofkprime})   &  SDSBR          & 0.0  & 0.0 & $1.2\times 10^{-3}$ & $9 \times 10^{-4}$ & idem & idem \\
   $r_{\mathrm{cu}}$           & (\ref{Sds_cu1})    &  SDSBRF1        & 0.0 & 0.0 &  0.0 & 0.5 & idem & idem \\
  $2*C_{\mathrm{cu}}$          &  (\ref{Sds_cu1})   & SDSC3           & 0.0 & 0.0 &  0.0 & -2.0 & -0.8 &  idem \\
  ${\mathrm{s_B}}$              &  (\ref{defBofkprime})   & SDSCOS           & 0.0 & 0.0 &  0.0 & 2.0 & idem & idem \\
  $\delta_d$                   &  (\ref{Sds_all})   &  SDSDC6         & 0.0  & 0.0 &  1.0 & 0.3&  0.3 & 0.0 \\
  $M_0$                        &     & SDSBM0          & 0.0 & 0.0 &  1.0 & idem &  idem & idem \\
  $M_1$                        &     & SDSBM1          & 0.0 & 0.0 &   0.2428 & idem & idem & idem\\
  $M_2$                        &     & SDSBM2          & 0.0 & 0.0 &  1.9995  & idem & idem & idem\\
  $M_3$                        &     & SDSBM3          & 0.0 & 0.0 &  -2.5709 & idem & idem & idem\\
  $M_4$                        &     & SDSBM4          & 0.0 & 0.0 &  1.3286  & idem & idem & idem\\
 \hline 
\end{tabular} \end{center}
\caption{Dissipation parameter as implemented in version 3.14-SHOM of the
WAVEWATCH III code, and values used in the tests presented here. In WWATCH, all parameters are accessible via the SDS3 namelist. The only parameters not defined in the present paper are $C_{\mathrm{lf}}$ and $C_{\mathrm{hf}}$ which act like switches to activate the BAJ parameterization for the part of the spectrum with saturation below and abovce the spectrum, respectively. Most of these are also included in version 3.14, except for $C_{\mathrm{cu}}$ which is needed for the TEST437 and TEST441 with results described here. The TEST405 can be ran with version 3.14.
The parameter $M_0$ is a switch for the correction or not of $B_r$ into $B'_r$, when $M_0=1$, as is the case here, the correction is not applied.} \label{tab:WAM4_parSDS}
\end{table*}

All parameters defining the dissipation source function and their numerical values are listed in table 3 for the wind-wave interaction term $S_{\mathrm{atm}}$ and table 4 for the 
wave-ocean interaction term $S_{\mathrm{oc}}$. We also recall that the nonlinear coupling coefficient (variable NLPROP in WWATCH) is set to  $2.78\times10^7$ in all cases, except for the two parameterizations mostly used here, with  $C_{\mathrm{nl}}=2.5\times10^7$ in TEST437 and TEST441. Although the best performance for most parameters is obtained with the TEST441 settings, its underestimation of extreme sea states may be a problem in some applications for which the TEST437 may be preferred. A full tuning of the model has not been tried yet and it is possible that a simple adjustment of $\beta_{\max}$ $C_{\mathrm{cu}}$, $r_{\mathrm{cu}}$ and $s_u$ may produce even better results. Finally, these parameters have been mostly adjusted for deep water conditions using ECMWF winds. Using other sources of winds for large scale applications may require a retuning of the wind source function, which can be best performed by a readjustment of  $\beta_{\mathrm{max}}$ . 

\end{appendix}

\ifthenelse{\boolean{dc}}
{}
{\clearpage}

\end{document}